\documentstyle[preprint,aps]{revtex}

\begin{document}

\draft

\title{
Analysis of the (N,xN') reactions
by quantum molecular dynamics
plus statistical decay model
}

\author{Koji Niita}
\address{
Advanced Science Research Center, \\
Japan Atomic Energy Research Institute,
Tokai, Ibaraki 319-11 Japan\\
and\\
Research Organization for Information Science and Technology,
Tokai, Ibaraki 319-11 Japan\\
}
\author{
Satoshi Chiba, Toshiki Maruyama, Tomoyuki Maruyama, Hiroshi Takada,\\
Tokio Fukahori, Yasuaki Nakahara, and Akira Iwamoto
}
\address{
Advanced Science Research Center,\\
Japan Atomic Energy Research Institute,
Tokai, Ibaraki 319-11 Japan
}
\maketitle
%
\begin{abstract}
We propose a model based on
quantum molecular dynamics (QMD) incorporated with
statistical decay model (SDM) to describe
various nuclear reactions in an unified way.
In this first part of the work, the basic ingredients of the
model are defined and the model is applied systematically to
the nucleon($N$)-induced reactions.
It has been found that our model can give
a remarkable agreement in
the energy-angle double differential cross sections of
$(N,xN')$ type reactions
for incident energies from 100 MeV to 3 GeV
with a fixed parameter set.
An unified description
of the major three reaction mechanisms of $(N,xN')$ reactions,
i.e. compound, pre-equilibrium and spallation processes,
is given with our model.
\end{abstract}
%
%
\section{Introduction}
Nuclear reactions reveal various aspects of hadronic many-body
problem as a function of
the target and projectile combination, the incident energy
and the angular momentum involved.
In nucleon induced reactions, for example, the compound process
is dominant in low energy region, while the pre-equilibrium
and spallation processes become more likely as the incident energy
increases.
In heavy-ion collisions, we also have to introduce various models
of different natures depending on each specific process.
However,
most of them are restricted to the specific energy regime or
specific phenomenon and some of them have too many parameters
to obtain a definite physical conclusion from the analysis.

The main purpose of the series of our work is to develop a model
which can describe the various aspects of nuclear reactions
in an unified way.
We try to seek a model with minimum number of parameters,
a wide range of applicability,
and a quantitative agreement with as many observables as
possible.
In addition to these requests, we demand the model to be
so simple that one can run its computer code on
work stations.

In the heavy-ion physics, microscopic models, which describe
reactions in terms of the dynamics of the interacting
nucleons, are commonly used to extract the information of the
nuclear matter under extreme conditions
from the final observables.
The most popular models of this type are the
Boltzmann-Uehling-Uhlenbeck/Vlasov-Uehling-Uhlenbeck
(BUU/VUU) \cite{vuu00},
the quantum molecular dynamics (QMD) \cite{qmd00},
and CASCADE type models \cite{cascade00,cascade01}.
So far these microscopic models have shedded light on
several exciting topics in heavy-ion physics, e.g. the
multifragmentation, the flow of the nuclear matter, and
the energetic particle production.
However
the parameters of the models such as
the effective interaction, elastic and
inelastic channels of $NN$ cross section, differ substantially
from one model to the other even in the same type of model.
Furthermore, these models have not been tested intensively
in much simpler light-ion reactions
except for an analysis of (p,xn) reaction
carried out by Peilert et al \cite{peilert}.
In their analysis, however, the lower part of neutron energy spectra
cannot be treated, since
statistical decay following the QMD process was not included.
We thus start the series of our work
from the analysis of the simplest type
of the reactions, the $(N,xN')$ (nucleon in, nucleon out)
reaction in this paper,
aiming to establish an unified model for various nuclear reactions.
In the subsequent works, we are planning to analyze $(N,x\pi)$,
$(\pi,xN)$ reactions and heavy-ion reactions.

We restrict our subject to the reactions
of nucleon-nucleus, meson-nucleus and nucleus-nucleus collisions
with energies well above Coulomb barrier up to several GeV/nucleon,
where the classical treatment of the collisions is justified
in a first-order approximation.
We do not deal with the phenomena which are dominated by the
quantum effects.
In this energy regime, the whole reaction process could be divided
into two parts, i.e. dynamical process and statistical process.
These two processes are well separated in their time scales.
In the dynamical process, the direct reactions, nonequilibrium
reactions, and dynamical formation of highly-excited fragments
take place during typical collision times of the order $10^{-22}$
sec. After that, the evaporation and fission decay, which we call
the statistical process, occur in the longer time scale
of the order $10^{-21}-10^{-15}$ sec.
We thus employ a two step model,
namely, we incorporate quantum molecular dynamics (QMD)
for the dynamical process with statistical decay model (SDM).
Similar hybrid models have been used in the analysis of the
heavy-ion collisions \cite{maru92a,g-qmd,f-qmd}.

In this paper, we define the basic ingredients of
QMD plus SDM model and discuss
how these two are combined without introducing
any additional parameter.
We then apply this model systematically
to $(N,xN')$ reactions,
and discuss which element in the model is crucial
for describing these reactions
and what is necessary to develop the model further.
In Sec. II we describe the details of the QMD,
the effective interaction, $NN$ elastic and inelastic cross
sections, the relativistic corrections and the statistical
decay model employed in our model.
In Sec. III we compare the various double differential cross
sections calculated by this model
with the experimental data for proton induced reactions
with incident energies from 100 MeV up to 3 GeV.
We summarize and conclude this work in Sec. IV.

\section{Description of the Basic Model}

As we mentioned above, our basic model consists of two part,
the quantum molecular dynamics (QMD) and
the statistical decay model (SDM).
The reason of employing the QMD model for the description of the
dynamical processes is that the QMD can calculate the fragment
formation in a natural and practical way.
Though the QMD method is widely used in the study of nuclear
fragmentation \cite{qmd00}, the details of the prescription
differ from author to author.
Aiming to establish a simple standard model,
we will start from the standard type of QMD,
taking into account of the relativistic
kinematics and the relativistic correction for the effective
interaction.
Additionally, we treat the resonances of nucleon, $\Delta$ and
$N^*(1440)$, and real pions with their isospin degrees of
freedom in the equation of motion.
For the statistical decay process, we use a simple prescription
including only the light particle evaporation.

\subsection{Quantum Molecular Dynamics}

\subsubsection{Basic equations and effective interaction}

The QMD method is a semi-classical simulation method in which
each nucleon state is represented by a Gaussian wave function of
width $L$,
\begin{equation}
\phi_i({\bf r}) = \frac{1}{(2\pi L)^{3/4}} \exp \left[
                - \frac{({\bf r} - {\bf R}_i)^2}{4L} +
                  \frac{i}{\hbar} {\bf r} \cdot {\bf P}_i \right],
\end{equation}
where ${\bf R}_i$ and ${\bf P}_i$ are the centers of position
and momentum of $i$-th nucleon, respectively.
The total wave function is assumed to be a direct product of
these wave functions.
Thus the one-body distribution function is obtained by the  Wigner
transform of the wave function,
\begin{equation}
f({\bf r},{\bf p}) =  \sum_i { f_i({\bf r},{\bf p}) },
\label{f0}
\end{equation}
\begin{equation}
f_i({\bf r},{\bf p})  =  8 \cdot
                        \exp\left[-{({\bf r}-{\bf R}_i)^2\over 2L}
                      -{2L({\bf p}-{\bf P}_i)^2\over \hbar^2}\right].
\label{fi}
\end{equation}
The time evolution of ${\bf R}_i$ and ${\bf P}_i$ is described by
Newtonian equations and the stochastic two-body collision term.
The Newtonian equations are derived on the basis of the
time-dependent variational principle \cite{qmd00} as
\begin{equation}
\dot{{\bf R}}_i =   \frac{\partial H}{\partial {\bf P}_i},
\;\;\;\;
\dot{{\bf P}}_i = - \frac{\partial H}{\partial {\bf R}_i},
\label{newton00}
\end{equation}
where the Hamiltonian $H$ consists of the single-particle
energy including the mass term
and the energy of the two-body interaction.
As for the effective interaction, we adopt the Skyrme type,
the Coulomb, and the symmetry terms  in this paper.
By using the Gaussian function of nucleons [Eq.\ (\ref{fi})],
we get
\begin{eqnarray}
H & = & \sum_i\;{ E_i } \nonumber \\
  &   & \; + \; {1\over 2}{A \over\rho_{0}}\sum_i<\rho_i>
        \; + \; {1\over 1+\tau}{B \over \rho_{0}^{\tau}}
                \sum_i<\rho_i>^{\tau} \nonumber \\
  &   & \; + \; {1\over 2}\sum_{i , j(\neq i)} c_{i} \, c_{j} \,
                {e^2\over|{\bf R}_i-{\bf R}_j|} \,
             \> {\rm erf}\left( |{\bf R}_i-{\bf R}_j|/\sqrt{4L}
                \right) \nonumber \\
  &   & \; + \; {C_{\rm s}\over 2\rho_0} \sum_{i , j(\neq i)} \,
                ( 1 - 2 | c_i - c_j | ) \; \rho_{ij},
\label{ham1}
\end{eqnarray}
with
\begin{equation}
E_i  =  \sqrt{ m_i^2 + {\bf P}_i^2} \;,
\end{equation}
where erf denotes the error function.
In the above equation, $c_i$ is 1 for protons and 0 for neutrons,
while $<\rho_i>$  is an overlap of density with
other nucleons defined as
\begin{eqnarray}
<\rho_i>     & \equiv & \sum_{j\neq i} \; \rho_{ij} \;
              \equiv  \sum_{j\neq i}
                     { \int { d{\bf r} \; \rho_i({\bf r}) \;
                       \rho_j({\bf r}) }} \nonumber \\
             & = & \sum_{j\neq i}{ (4\pi L)^{-3/2}
                  \exp \left[ - ( {\bf R}_i - {\bf R}_j ) ^2
                  / 4L \right] },
\label{rhoij}
\end{eqnarray}
with
\begin{eqnarray}
\rho_i({\bf r}) & \equiv & \int \frac{d{\bf p}}{(2\pi \hbar)^3}
                         \; f_i({\bf r},{\bf p}) \nonumber \\
                & = &  (2\pi L)^{-3/2}
                     \exp \left[ - ( {\bf r} - {\bf R}_i ) ^2
                     / 2L \right].
\end{eqnarray}
In this paper we use the parameters $A=-219.4$ MeV,
$B=165.3$ MeV, and $\tau=4/3$ which yield a compressibility of
$K=237.3$ MeV, saturation at $\rho = \rho_0 = 0.168
{\rm fm}^{-3}$ and
a binding energy of 16 MeV per nucleon for infinite nuclear
matter.
The symmetry energy parameter $C_{\rm s}$ is chosen to be 25 MeV.
The width of Gaussian $L$ is a parameter of the QMD and fixed
as $L=2.0 \; {\rm fm}^2$ in this paper.

\subsubsection{Two-body collision term}

In addition to the Newtonian equation Eq.\ (\ref{newton00}),
the time evolution of the system is affected by the
two-body collision term.
In the QMD method, the stochastic two-body collision process
is introduced in a phenomenological way on the analogy of
the test-particle calculation of the BUU collision term
\cite{vuu00}.
It includes the Pauli blocking factor
$(1-f({\bf r},{\bf p},t))$,
which is lacking in the cascade collision process
\cite{cascade00,cascade01}.
We follow basically the prescription of the two-body collision term
used in the BUU calculation done by Wolf et al. \cite{wolf,wolf2},
and modify it to extend the energy range up to 3 GeV.
We thus describe here the outline of the procedure of
Ref.\ \cite{wolf,wolf2}
and explain the extensions introduced in this paper.
Further details of the collision term and
the dynamics of $\Delta$'s, $N^*$'s, and pions discussed
below can be found in Ref.\ \cite{wolf,wolf2}.

It is assumed that two particles collide if their impact parameter
defined in a covariant way is smaller than a given value
$b_{{\rm max}}=\sqrt{\sigma/\pi}$ obtained
from cross section~$\sigma$.
The collisions are considered as instantaneous interaction and
a collision event is specified by the two points in space-time
where the collision happens.
Therefore it is hard to retain the covariance,
since one has to choose a common reference frame
for the QMD calculations.
Hence the average proper time of the collision
points defined by each particle is used to determine
the time step in which the collision happens.
This collision prescription was checked for
heavy-ion collisions from 400 MeV/nucleon to 2.1 GeV/nucleon,
and it was found that the disturbance of the covariance
was very small \cite{wolf}.

In order to treat the reactions with high bombarding energies,
we include in our QMD simulation the nucleons ($N$), deltas
($\Delta$(1232)), $N^*$(1440)'s and pions with their
isospin degree of freedom.
The $\Delta$'s and $N^*$'s are propagated in the same
interactions as the nucleons except for the symmetry term,
while pions feel only the Coulomb interaction.
The creation and absorption of these particles are
treated in the collision term.
In the following, we list all channels included in the collision
term, where $B$ denotes a baryon and $N$, more specifically,
a nucleon:
\begin{eqnarray}
\begin{array}{llllllll}
1.\;& B_i   & +\;& B_j    & \to \;& B_i  & +\;& B_j  \\
2. &  N     & +  & N      & \to & N      & +  & \Delta  \\
3. &  N     & +  & \Delta & \to & N      & +  & N \\
4. &  N     & +  & N      & \to & N      & +  & N^*  \\
5. &  N     & +  & N^*    & \to & N      & +  & N \\
6. &  N     & +  & \pi    & \to & \Delta &    & \\
7. &  N     & +  & \pi    & \to & N^*    &    & \\
8. & \Delta & +  & \pi    & \to & N^*    & .  & \\
\end{array}
\label{channel1}
\end{eqnarray}
The channel 8 has been added to the prescription of
Wolf et al. \cite{wolf},
which is the inverse process of the additional decay channel
of $N^*$(1440)'s (cf. channel 11 below).

We use the following parametrization
for all baryon-baryon elastic cross section
(channel 1 in Eq.\ (\ref{channel1})),
\begin{equation}
\sigma =
\frac{C_1}{1+100 \sqrt{s} \; '} + C_2 \; ({\rm mb}),
\label{signn1}
\end{equation}
with
\begin{equation}
\sqrt{s} \; ' =
         \max( 0, \sqrt{s} - M_i - M_j - {\rm cutoff} ) \;
         ({\rm GeV}),
\end{equation}
where cutoff is 0.02 GeV for nucleon-nucleon channel, while
it is zero for the others.
This is the conventional
Cugnon parametrization form \cite{vuu00,cascade01}.
We use this form up to
$\sqrt{s} \; ' = 0.4286 \; ({\rm GeV})$,
which corresponds to 1 GeV Lab energy
for nucleon-nucleon case.
Above 1 GeV, we parametrize the experimental
$p$-$p$ and $p$-$n$ elastic
cross section \cite{ptable,data1} as,
\begin{eqnarray}
\sigma =
C_3 \left[ 1 - \frac{2}{\pi} \tan^{-1} \left(
1.5 \sqrt{s} \; ' - 0.8 \right) \right]
+ C_4 \; ({\rm mb}).
\label{signn2}
\end{eqnarray}
In order to connect Eqs.\ (\ref{signn1}) and (\ref{signn2})
smoothly, we slightly modified the parameters of Cugnon
\cite{vuu00,cascade01}.
The actual values of the parameters $C_i$ in the above equations
are listed in Table \ref{table1}.
The angular distribution of the elastic channels is taken from
the same form as Cugnon parametrization \cite{vuu00,cascade01}.

For the production of baryonic resonances
(channel 2 and 4 in Eq.\ (\ref{channel1})),
we adopt the total cross section based on the
parametrization of VerWest and Arndt
\cite{west}, in which the pion cross sections are parametrized
assuming the pions are produced through baryonic resonances.
Their parametrization was performed by fitting the
experimental data up to 1.5 GeV incident energy.
In order to extend the energy range up to 3 GeV,
we have modified the parameters in the following way.
In the model of VerWest and Arndt,
the cross sections are parametrized according
to the initial and final total isospin $i$ and $f$ of the
two-nucleon system \cite{west} as
\begin{equation}
\sigma_{if}(s) = \frac{\pi(\hbar c)^2}{2p^2}
\; \alpha \left( \frac{p_r}{p_0}\right)^{\beta}
\frac{m_0^2 \Gamma^2 (q/q_0)^3}
{(s^* - m^2_0)^2 + m_0^2 \Gamma^2},
\label{wests}
\end{equation}
where
\begin{eqnarray}
p_0^2 & = & {1\over 4} s_0 - m^2_N , \;\;\;\;\;
s_0 = (m_N+m_0)^2 , \nonumber \\
p_r^2(s) & = &
\frac{[s-(m_N-\langle M \rangle )^2][s-(m_N+\langle M \rangle )^2]}
{4s}, \\
q^2(s^*) & = &
\frac{[s^*-(m_N-m_\pi )^2][s^*-(m_N+m_\pi )^2]}
{4s^*}, \nonumber \\
s^* & = & \langle M \rangle^2 , \;\;\;\;
q_0 = q(m_0^2), \nonumber
\end{eqnarray}
and $\langle M \rangle$ is the mean mass of the resonance \cite{west}
obtained from a Breit-Wigner distribution with $M_0 = 1220$ MeV,
$\Gamma_0 = 120$ MeV for the $\Delta$ and $M_0 = 1430$ MeV
and $\Gamma_0 = 200$ MeV for the $N^*$.
 From these cross sections, we determine the production cross
section of $\Delta$'s \cite{wolf} as
\begin{eqnarray}
\begin{array}{lllllllll}
p & + & p & \to & n & + & \Delta^{++}
&     :     & \sigma_{10}+{1\over 2}\sigma_{11} ,\\[0.5ex]
p & + & p & \to & p & + & \Delta^{+}
&     :     & {3\over 2}\sigma_{11} ,\\[0.5ex]
n & + & p & \to & p & + & \Delta^{0}
&     :     & {1\over 2}\sigma_{11}+{1\over 4}\sigma_{10} ,\\[0.5ex]
n & + & p & \to & n & + & \Delta^{+}
&     :     & {1\over 2}\sigma_{11}+{1\over 4}\sigma_{10} ,\\[0.5ex]
n & + & n & \to & p & + & \Delta^{-}
&     :     & \sigma_{10}+{1\over 2}\sigma_{11} ,\\[0.5ex]
n & + & n & \to & n & + & \Delta^{0}
&     :     & {3\over 2}\sigma_{11} .
\end{array}
\label{delta}
\end{eqnarray}
We have effectively included the cross section of the $\pi d$ final
state, parametrized as $\sigma^d_{10}$ in \cite{west},
in the cross section of $\sigma_{10}$.

We assume in this paper that the cross section $\sigma_{01}$
in \cite{west} contributes only to the $N^*$ production
independently
of the isospin components.
Thus we rename $\sigma_{01}$ as $\sigma_{N^*}$, and the
$N^*$ production cross sections are given by
\begin{eqnarray}
\begin{array}{lllllllll}
p & + & p & \to & p & + &  N^{*+}
&     :     & {3\over 2}\sigma_{N^*} ,\\[0.5ex]
n & + & p & \to & p & + & N^{*0}
&     :     & {3\over 4}\sigma_{N^*} ,\\[0.5ex]
n & + & p & \to & n & + & N^{*+}
&     :     & {3\over 4}\sigma_{N^*} ,\\[0.5ex]
n & + & n & \to & n & + & N^{*0}
&     :     & {3\over 2}\sigma_{ N^*} .\\[0.5ex]
\end{array}
\label{nstar}
\end{eqnarray}
The new parameters in Eq.\ (\ref{wests}) are given in
Table \ref{table3}.
In order to determine these parameters and the parameters of
elastic cross section in high energy part defined in
Eq.~(\ref{signn2}),
we fitted the experimental $p$-$p$ and $p$-$n$ cross sections
\cite{ptable,data1}.
In Fig.\ \ref{pptot}, we show the $p$-$p$ (a) and $p$-$n$ (b)
total (solid line), elastic (long dashed line),
and inelastic (dot-dashed line) cross sections.
The inelastic cross section is the sum of the
$\Delta$ (short dashed line) and $N^*$
(dotted line) production cross section, calculated by
Eqs.~(\ref{signn1},\ref{signn2},\ref{wests},\ref{delta},\ref{nstar}).
In the same figure, we show the corresponding
experimental total (open circles),
elastic (open triangles), and inelastic (open boxes)
cross sections \cite{data1} with error bars.
For the $p$-$p$ case, the present parametrization of
elastic and inelastic cross sections fits to the data
for whole energy range up to 3 GeV except for the some
deviation around 1 GeV, which is due to the
elastic cross section and does not affect the result.
On the other hand, for $p$-$n$ case where only the total cross
section is available in the data,
we fitted it at the energy higher than 0.7 GeV up to 3 GeV,
and adopt the Cugnon's type elastic cross section
in the low momentum region instead of the free elastic cross
section.

In Fig.\ \ref{pppi} we show the pion cross section of
$pn \to nn\pi^+ + pp\pi^-$ (solid line)
obtained by our new parametrization
of Table \ref{table3}.
In the same figure, the gray bold line denotes the result of the
original
parametrization of VerWest and Arndt \cite{west},
while the experimental data are taken from Ref.\ \cite{data1}.
By this new parameter set,
our pion production cross section below 1.5 GeV
does not differ so much from
the original results of VerWest and Arndt \cite{west},
which are essentially the same as the data.
However, above 1.5 GeV,
our result fitted the experimental data,
while the result obtained by extrapolating the
original parametrization of VerWest and Arndt to higher energy
shows a big bump, which has no experimental support.

In the higher energy region, the role of $N^*$ becomes important.
One of the good quantities which shows the characteristics of the
higher resonances is the elementary two pion production cross
section.
In the present prescription it is described only by $\sigma_{N^*}$
combined with the decay modes of the resonances which will be
mentioned below.
For example, it is shown in our prescription that
\begin{eqnarray}
\sigma(pn \to pp \pi^0 \pi^-) & = & \frac{1}{15} \sigma_{N^*},
\nonumber \\
\sigma(pn \to pn \pi^+ \pi^-) & = & \frac{5}{12} \sigma_{N^*},
\label{twopi} \\
\sigma(pp \to pp \pi^+ \pi^-) & = & \frac{1}{3} \sigma_{N^*}.
\nonumber
\end{eqnarray}
We thus plot the $\sigma_{N^*}$ in Fig.\ \ref{signn3}
as well as the experimental two pion production cross sections
\cite{data1} scaled by above factors.
This figure shows that our parametrization fitted the gross
features of the experimental data for the energy range up to
3 GeV.
Although this parametrization should be modified if two delta
production channel or the direct two pion decay of $N^{*}$ or
higher resonances are included,
the present prescription of the elementary inelastic channels
could roughly reproduce the experimental single and two
pion production cross sections for the energy range up to
3 GeV.

We do not take into account the direct s-state pion production
mechanism but all pions are assumed to be produced through
baryonic resonances.
The masses of the resonances are randomly distributed according to
the Breit-Wigner distribution with the momentum-dependent width
\cite{moniz}, i.e.,
\begin{equation}
f(M) = \frac{0.25 \Gamma^2}{(M-M_r)^2 + 0.25 \Gamma^2},
\label{masdis1}
\end{equation}
with
\begin{equation}
\Gamma = \left( \frac{q}{q_r} \right)^3 \frac{M_r}{M}
         \left( \frac{v(q)}{v(q_r)} \right)^2 \Gamma_r,
\label{masdis2}
\end{equation}
where $q$ denotes the c.m. momentum in the $\pi N$ channel,
and index $r$ refers to the values at the mass $M_r$, and
\begin{equation}
v(q) = \frac{\beta_r^2}{\beta_r^2 + q^2}.
\label{masdis3}
\end{equation}
We have applied this momentum dependent width not only to
the $\Delta$-resonance but also to $N^*$(1440). The values
of the parameters used in this paper are listed
in Table~\ref{table4}.

Another important ingredient of the resonance production
(channel 2 and 4 in Eq.\ (\ref{channel1})) is the angular
distribution of the resonances.
Wolf et al. \cite{wolf} parametrized the angular distribution
of the experimental data \cite{ange1} for
$p+p \to  n+\Delta^{++}$ and assumed the same angular dependence
for each isospin channel in the following form,
\begin{equation}
g_{\rm R}(s,\cos\theta) = a_0(s)
\; [ a_1(s) + 3 a_3(s) \cos^2\theta ],
\label{angwolf1}
\end{equation}
with
\begin{equation}
a_0(s) = \frac{1}{4 \pi ( a_1(s) + a_3(s) ) }.
\label{angwolf2}
\end{equation}
The values of $a_1(s)$ and $a_3(s)$ are given in Table~\ref{table5}.
In the high energy region $\sqrt{s} > 2.4 $ GeV, which
corresponds to the laboratory energy higher than 1.2 GeV,
this angular distribution is assumed to be constant,
since there is no experimental data to be fitted in this energy
region.
However, above the resonance region $E_{{\rm lab}} > 1.2 $~GeV,
this is not justified because, for example,
the angular distribution of protons from
$^{27}$Al$(p,p')$ at 3.17 GeV reaction
calculated by Eq.\ (\ref{angwolf1})
deviates from the experimental data
(cf. discussions on Fig.\ \ref{enyo1} in the next section).

In order to get a better parametrization for high energy part,
we assume that the angular dependence is effectively
written as a sum of $g_{\rm R}$ and another term $g_{\rm D}$ as
\begin{equation}
g_{\rm A}(s,\cos\theta) = \frac{1}{2} \, \left[
                    g_{\rm R}(s,\cos\theta)
                  + g_{\rm D}(s,\cos\theta) \right].
\label{angtot}
\end{equation}
where
\begin{equation}
g_{\rm D}(s,\cos\theta) = b_0(s)
                    \; \exp \left[ - 2 \, p^2(s) \, b_1(s)
                    \; ( 1 - \cos\theta )  \right],
\label{angdir1}
\end{equation}
and
\begin{eqnarray}
b_0(s) & = & \frac{p^2(s) \, b_1(s)}
           {\pi \, ( \, 1- \exp[ -4 \, p^2(s) \, b_1(s)] \, ) },
            \\[3ex]
b_1(s) & = & \frac{0.14 \, s^2 \, [3.65 \,
           ( \sqrt{s} -m_N -m_R) \, ]^6 }
           {1 + [3.65 \, ( \sqrt{s} -m_N -m_R) \, ]^6 }, \\[3ex]
p^2(s) & = & \frac{[s-(m_N-m_R)^2][s-(m_N+m_R)^2]}
           {4s}.
\end{eqnarray}
This form of $g_{\rm D}$ is obtained by modifying the
Cugnon parametrization \cite{vuu00,cascade01}
of the $NN$ elastic angular distribution so as to
trace the angular distribution of Eq.~(\ref{angwolf1})
in the resonance region, and approach the elastic-like angular
distribution for the higher momentum region.
We use this angular distribution for $\Delta$-resonance and
$N^*$(1440).
For the latter case, we apply this formula by shifting
the energy $\sqrt{s}\;$ by the mass
difference of the two resonances, i.e., 208 MeV.
The energy dependence of this angular distribution is shown
in Fig.\ \ref{angf}, where we plot
$g_{\rm R}$ (gray bold dashed lines)
and $g_{\rm D}$ (solid lines) in (a), while
$g_{\rm R}$ (gray bold dashed lines)
and $g_{\rm A}$ (solid lines) in (b).
In these figures, we symmetrized the elastic-like angular
distribution,
i.e., $ \frac{1}{2} [ g_{\rm D}(s,\cos\theta)
+ g_{\rm D}(s,-\cos\theta)] $,
in order to compare it with the angular distribution
of Eq.\ (\ref{angwolf1}) (gray bold dashed lines).
In the next section, we will discuss
the dependence of the final results of $(N,xN')$ reactions on this
angular distribution of the hadronic resonances.

The cross sections for the channel 3 and 5 in Eq.\ (\ref{channel1})
are determined by the law of detailed balance from the cross
section of channel 2 and 4 with taking into account the mass
dependence of the cross section \cite{wolf2}.

For the pion absorption cross section on nucleons (channel 6 and 7
in Eq.\ (\ref{channel1})) we take the maximum cross sections from
the particle data table \cite{ptable} and scale them according to
the Breit-Wigner formula.
For the case of pion absorption on $\Delta$
(channel 8 in Eq.\ (\ref{channel1})),
we assume the same cross section as the channel 7 with shifting
the energy by the mass difference of the $\Delta$ and nucleon.

Apart from the collision term, we take into account the decay of
the baryonic resonances during the propagation as
\begin{eqnarray}
\begin{array}{rlllll}
9.  & \Delta & \to & N      & +  & \pi  \\
10. &  N^*   & \to & N      & +  & \pi  \\
11. &  N^*   & \to & \Delta & +  & \pi  \\
\end{array}
\label{channel2}
\end{eqnarray}
The decay probability of the resonances is determined by
a exponential decay law using their momentum-dependent width
Eq.\ (\ref{masdis2})
and their proper time.
The decay is assumed to be isotropic in their rest frame.
The branching ratio of the channel 10.\ and 11.\
of Eq.\ (\ref{channel2})
is taken from the particle data table \cite{ptable} as
$\Gamma_{N^* \to \Delta+\pi}
/ [\Gamma_{N^*\to \Delta+\pi} + \Gamma_{N^*\to N+\pi}]
= 0.4$.
The other branching ratios concerning their isospin are determined
from the appropriate Clebsch-Gordan coefficients.

\subsubsection{Relativistic corrections}

Non-covariant framework is
another problem of applying the QMD method
to the reactions at higher bombarding energies.
As explained above, we have already introduced the relativistic
form of energy expression in the Hamiltonian
Eq.\ (\ref{ham1}) and adopted the relativistic kinematics in the
collision term.
However a covariant formulation of the interaction term is necessary
for a full relativistic description.
A Lorentz covariant extension of the QMD, dubbed relativistic
quantum molecular dynamics (RQMD), has been proposed
by Sorge et al. \cite{rqmd}
based on the Poincar\'e-invariant constrained Hamiltonian dynamics.
Although the RQMD is a numerically feasible extension of QMD
toward a fully covariant approach, it still costs too much computing
time to apply the RQMD model to heavy systems.
We thus make the following alternative extension of QMD
and include the main part of the relativistic dynamical effects
in our model.

Lehmann et al. \cite{lehmann} have compared
the time evolution of the phase space and the particle production
obtained by QMD and RQMD, looking for the relativistic effects
in heavy-ion collisions in the intermediate energy regime.
They found that there is no significant difference between the
results of QMD and RQMD in the $\eta$ and $\pi$ meson production
cross sections \cite{lehmann} and the proton inclusive spectra
\cite{rqmdt}.
The difference appeared only in the values of the maximum density
\cite{rqmdt} and the transverse flow \cite{rqmdm}.
Both are larger in RQMD.
Their studies showed that a large part of the difference comes
from the Lorentz contraction of the initial phase space
distribution in RQMD.
If this Lorentz contraction is employed in the normal QMD, however,
the transverse flow is overestimated.
Thus they concluded that in RQMD this effect is partially
counterbalanced by the covariant treatment of the interaction,
but there still remains an increased flow
compared with the normal QMD calculation.
Based on their investigation, we introduce in this paper
the Lorentz-scalar quantities into the
arguments of the interactions in Eq.\ (\ref{ham1})
as well as
the Lorentz contraction of the initial phase space distribution.
By these modifications the main part of the
relativistic dynamical effects would be approximately
included in our QMD.

All arguments of the interaction in Eq.\ (\ref{ham1}) are written
as a function of the squared spatial distance:
\begin{equation}
{\bf R}_{ij}^2 = ({\bf R}_i - {\bf R}_j)^2.
\end{equation}
In the RQMD \cite{lehmann,rqmdt},
these arguments are replaced by the squared transverse
four-dimensional distance $-q_{{\rm T}_{ij}}^2$ defined as
\begin{equation}
-q_{{\rm T}_{ij}}^2 = -q_{ij}^2 + \frac{(q_{ij} \cdot p_{ij})^2}
                                 {p_{ij}^2},
\end{equation}
where $q_{ij}$ is the four dimensional distance
$q_i-q_j$, while $p_{ij}$ denotes the sum of the four momenta of
the two particle $p_i+p_j$.
In the c.m.s of the particle $i$ and $j$,
the squared covariant transverse distance $-q_{{\rm T}_{ij}}^2$
reduces to the usual squared distance.
We therefore change the argument in Eq.\ (\ref{ham1}) from
the ${\bf R}_{ij}^2$ to the squared distance in the c.m.s.
of the two particles
$\tilde {\bf R}_{ij}^2$,
where the tilde means the quantities defined
in the c.m.s of the two particles,
\begin{equation}
 \tilde {\bf R}_{ij}^2  =  {\bf R}_{ij}^2 + \gamma_{ij}^2
             \left( {\bf R}_{ij} \cdot {\vec \beta}_{ij} \right)^2,
\end{equation}
with
\begin{equation}
{\vec \beta}_{ij} = \frac{{\bf P}_i + {\bf P}_j}
                         { E_i + E_j },  \;\;\;\;\;\;\;\;
\gamma_{ij} = {1\over \sqrt{1-{\vec \beta}_{ij}^2}}.
\end{equation}

By this change, all interactions of the Hamiltonian
[Eq.\ (\ref{ham1})] depend also on the momentum.
The form of the equation of motion [Eq.\ (\ref{newton00})]
changes to
\begin{eqnarray}
\dot{{\bf R}}_i & = & \frac{{\bf P}_i}
                           {\sqrt{m_i^2+{\bf P}_i^2}}
                    + \sum_j D_{ij} \,
                      \frac{\partial \;{\tilde {\bf R}_{ij}^2}}
                           {\partial \,{\bf P}_i},
\label{eom01} \\
\dot{{\bf P}}_i & = & - \sum_j D_{ij} \,
                      \frac{\partial \;{\tilde {\bf R}_{ij}^2}}
                           {\partial \,{\bf R}_i},
\label{eom02}
\end{eqnarray}
with
\begin{eqnarray}
D_{ij} & = &
     - \> \frac{1}{2} \frac{A}{\rho_0} \> \frac{1}{2L} \> \rho_{ij}
     \nonumber \\
 & & - \> \frac{1}{1+\tau} \frac{B}{\rho_0^{\tau}} \> \frac{\tau}{2}
     \left( < \rho_i>^{\tau-1} + < \rho_j>^{\tau-1}
     \right) \> \frac{1}{2L} \> \rho_{ij}      \nonumber \\
 & & + \> \frac{e^2}{2} c_i c_j \left\{ - \frac{1}
       {|\tilde {\bf R}_{ij}|}
     \> {\rm erf} \left( \frac{|\tilde {\bf R}_{ij}|}{\sqrt{4L}}
     \right) + 8 \pi L \> \rho_{ij} \right\}
     \frac{1}{\tilde {\bf R}_{ij}^2}
     \nonumber \\
 & & - \> \frac{C_{\rm s}}{2\rho_0} \>
     (1-2|c_i-c_j|) \> \frac{1}{2L} \> \rho_{ij},
\end{eqnarray}
and
\begin{eqnarray}
 \frac{\partial \; \tilde {\bf R}_{ij}^2}{\partial \, {\bf R}_i}
 & = & 2 {\bf R}_{ij} + 2 \gamma_{ij}^2 \left( {\bf R}_{ij}
         \cdot \vec \beta_{ij} \right)
         \vec \beta_{ij} \label{dev1} \label{eom11} \\
 \frac{\partial \; \tilde {\bf R}_{ij}^2}{\partial \, {\bf P}_i}
 & = & \frac{2 \gamma_{ij}^2}{E_i + E_j} \left( {\bf R}_{ij}
       \cdot \vec \beta_{ij} \right) \nonumber \\
 & \times &
     \left\{ {\bf R}_{ij} + \gamma_{ij}^2
     \left( {\bf R}_{ij} \cdot \vec \beta_{ij} \right)
     \left( \vec \beta_{ij} - \frac{ {\bf P}_i}{E_i} \right)
     \right\} \;, \label{eom12}
\end{eqnarray}
where $\rho_{ij}$ is defined in Eq.\ (\ref{rhoij}).

We also introduce the Lorentz-scalar quantities into the one-body
phase-space distribution function Eq.\ (\ref{fi}) as
\begin{equation}
f_{ij}  =   8 \cdot
              \exp\left[ - {1\over 2L} \> \tilde {\bf R}_{ij}^2
              -{2L\over \hbar^2} \> \tilde {\bf P}_{ij}^2 \right] ,
\end{equation}
where $\tilde {\bf P}_{ij}^2$ denotes the squared relative momentum
in the c.m.s of the particle $i$ and $j$,
which is expressed for two particles with the same mass
as
\begin{equation}
\tilde {\bf P}_{ij}^2 = {\bf P}_{ij}^2 - ( E_i - E_j )^2,
\end{equation}
with
\begin{equation}
{\bf P}_{ij} = {\bf P}_i - {\bf P}_j.
\end{equation}

At the starting point of the QMD calculation,
we boost the ground state of projectile
(and target as well if c.m.s of the
target and projectile is chosen as a reference frame)
according to the beam energy.
The coordinate ${\bf R}_{iz}^{\rm b}$ and momentum
${\bf P}_{iz}^{\rm b}$
of the nucleon in the beam direction $z$
after the boost are obtained by the Lorentz transformation
from those before the boost, ${\bf R}_{iz}$ and ${\bf P}_{iz}$,
as
\begin{eqnarray}
{\bf R}_{iz}^{\rm b} & = & ( {\bf R}_{iz} - {\bf R}_{0z} )
                      \> / \, \gamma \> + \> {\bf R}_{0z}, \\
{\bf P}_{iz}^{\rm b} & = & \gamma \> ( {\bf P}_{iz} + \beta E_i ),
\end{eqnarray}
where ${\bf R}_0$ denotes the initial c.m.\ coordinate of the
nucleus, while $\beta$ and $\gamma$ are the boosting velocity
and its gamma factor, respectively.
At this moment, the potential energy of the system
and the phase-space distribution function keep the same values
as those before the boost due to their Lorentz-scalar properties
discussed above.
During the propagation of the boosted nuclei, however,
those quantities are changing slightly even in the above
prescriptions, since the equations of motion Eq.\ (\ref{eom01})
and (\ref{eom02}) are not covariant.
But the disturbance due to the non-covariant feature of the
equation of motion is negligibly small up to the energy
3 GeV/nucleon.
We thus introduce the relativistic corrections discussed
above to the non-covariant QMD to save the computing time
instead of using  the full covariant framework.

We should mention here that
if one employs the Lorentz contraction for the boosted initial
state but does not replace the arguments of
the interaction and phase-space distribution by
the Lorentz-scalar ones,
the potential energy decreases about
80 MeV and the phase-space factor at each nucleon's point
changes about 40 \% after the boost
in the case of $^{40}{\rm Ca}$ even at 1~GeV/nucleon boosting energy.
This means that the boosted ground state obtained by this way
may decay spontaneously before it collides with the other nucleus.
By our prescription, we are free from this problem.

We have checked the above prescription
by analyzing the transverse flow,
which is sensitive to the treatment of the relativity
as discussed in Ref.\ \cite{lehmann}.
In Fig.\ \ref{fig-flow} we show the energy dependence of
so called directed transverse momentum $<{\bf P}^{{\rm dir}}_x>$,
which is a measure for the transverse flow and
defined by
\begin{equation}
<{\bf P}^{{\rm dir}}_x> = {1\over N} \sum_{i=1}^N {\rm sign} \left[
              Y_{{\rm CM}}(i) \right] {\bf P}_{ix},
\end{equation}
where $Y_{{\rm CM}}(i)$ is the rapidity of the $i$-th baryon in the
c.m.s and ${\bf P}_{ix}$ its transverse momentum in the
reaction plane.
We plot the result of the present QMD
simulation (solid line with full boxes) as a differences from
that of RQMD \cite{rqmdt}
for $^{40}{\rm Ca}$ + $^{40}{\rm Ca}$ reactions
at $b$ = 2~fm, for the energy range from 150 MeV/nucleon
to 4 GeV/nucleon.
In these calculations, we use the same ground states
(mentioned below), the same gaussian wave packets and
the same interaction (mentioned before).
We omitted the Coulomb interaction and two-body collision
term for a simplicity.
In Fig.\ \ref{fig-flow}, we also plot the other results obtained
by the standard QMD (dot-dashed line with open circles)
without the initial Lorentz contraction and without the
relativistic corrections,
and by the standard QMD only with the Lorentz contraction
(dashed line with open boxes).
As mentioned before, the Lorentz contraction of the initial
phase space distribution increases the flow, which is shown by
the change from the dot-dashed line to the dashed line.
By the full covariant treatment of the interaction, however,
this effect is counterbalanced, but still remains an increased flow
\cite{lehmann}.
As seen in Fig.\ \ref{fig-flow},
our prescription does not deviate so much from the full covariant
treatment up to 3 GeV/nucleon.
We thus expect that our QMD simulation with the relativistic
corrections is very close to the covariant simulation RQMD
in this energy regime.

At much higher energy, however, our result is decreasing linearly
from that of RQMD.
This deviation comes from the different treatment of the potential
between the RQMD \cite{rqmdt} and our QMD;
a Lorentz scalar type in the former, while a time-component of the
vector type in the later, respectively.
This is understood qualitatively by considering
a single particle motion under a fixed external potential $U$.
In the Lorentz scalar treatment of the potential $U$,
the single particle energy $p_i^0$ is expressed in this simple
case as
\begin{equation}
p_i^0 = \sqrt{p_i^2 + m_i^2 + 2m_i U}.
\end{equation}
Accordingly the equation of motion is
\begin{equation}
\dot{{\bf P}}_i = -
     \frac{m_i}{p^0_i} \frac{\partial U}{\partial {\bf R}_i}.
\end{equation}
On the other hand in our prescription they are
\begin{equation}
p_i^0 = \sqrt{p_i^2 + m_i^2 } + U,
\end{equation}
and
\begin{equation}
\dot{{\bf P}}_i = -  \frac{\partial U}{\partial {\bf R}_i}.
\end{equation}
In this test calculations, the form of
$\frac{\partial U}{\partial {\bf R}_i}$,
which is attractive in the beginning,
is almost the same in our QMD and in the RQMD.
Thus the force of our QMD becomes larger and deviates linearly
from that of the RQMD as energy increases.
Above 3 GeV/nucleon, therefore, the full covariant prescription
is necessary to describe the reactions particularly
the nucleus-nucleus collisions.
This is out of scope of this paper.

Some details of actual numerical calculations should be mentioned
here, since all potential terms depend on the momentum by the
relativistic corrections.
In order to keep a numerical accuracy, we use the second order
Runge-Kutta method to integrate the equations of motion.
For the energy conservation for the collision term, we assume
\begin{equation}
E_i + E_j + U_{{\rm pot}} = E_i' + E_j' +U_{{\rm pot}}',
\label{econ1}
\end{equation}
where $E_i$, $E_j$, and $E_i'$, $E_j'$ are the energies of particle
$i$ and $j$ before and after the collision, while $U_{{\rm pot}}$
and $U_{{\rm pot}}'$ the potential energy of the system.
We determine iteratively the final momenta of
the colliding particles so as to satisfy the energy conservation
Eq.\ (\ref{econ1}).
This prescription is applied to the channels 1 to 5 in
Eq.\ (\ref{channel1}), 9 to 11 in Eq.\ (\ref{channel2}).
For the pion absorption channels of 6, 7 and 8
in Eq.\ (\ref{channel1}),
the energy conservation is written as
\begin{equation}
E_i + E_j + U_{{\rm pot}} = E_R  +U_{{\rm pot}}',
\label{econ2}
\end{equation}
where $E_R$ is the resonance energy.
In this case, we determine iteratively the rest mass of the
resonance to conserve the energy.

\subsubsection{Properties of ``ground state''}

An important ingredient of the QMD calculation is how to
determine the initial phase space distribution
of the projectile and target.
For that we cannot use the real ground state (energy minimum state)
of the system defined by the Hamiltonian Eq.\ (\ref{ham1}),
since the model does not have the Fermionic properties.
However, it is necessary to obtain a stable ``ground state''.
Also some typical properties of the real ground
state should approximately be fulfilled,
especially the binding energy and
phase space distribution.
To get such ``ground state'', we employ the following
random packing procedure \cite{f-qmd};

We distribute the centers of position ${\bf R}_i$ of the individual
Gaussian wave packet
according to a distribution of the Woods-Saxon shape
with the radius $R_0 = 1.124 \; A^{1/3} - 0.5 $ fm
and the diffuseness parameter $a = 0.2 $ fm.
We cut off the Woods-Saxon tail at
$R_{{\rm max}} = 1.124 \; A^{1/3} $ fm.
In this procedure, we impose minimum distance between the centers
of the Gaussians in order to reduce the density fluctuation.
We use 1.5 fm for the identical nucleons and 1.0 fm for
the other.

Now we can calculate the density and potential energy at any point
(here, we do not need the relativistic correction
discussed in the previous sub-section).
Then the center of momentum ${\bf P}_i$ is randomly sampled
from the sphere of radius $p_{\rm F}({\bf R}_i)$ which is the Fermi
momentum obtained by the local Thomas-Fermi approximation.
This sampling is rejected and another value is sampled
if the sum of kinetic and potential energy of the particle is
positive or the phase space factor $f({\bf R}_j,{\bf P}_j)$
(cf. Eqs.~(\ref{f0}) and (\ref{fi}) )
for the nucleon $j$ which have been previously accepted
violates the Pauli principle \cite{f-qmd}.

Finally, we check the total binding energy with the
simple mass formula \cite{bm1}, i.e.,
\begin{eqnarray}
E_{{\rm bin}} = & - & 15.56 \; A
            +   17.23 \; A^{2/3} \nonumber \\
          & + & 46.57 \; \frac{(N-Z)^2}{2\,A}
            +   \frac{3}{5} \; \frac{Z^2e^2}{1.24\,A^{1/3}}.
\label{liquid-e}
\end{eqnarray}
If the binding energy per nucleon obtained by our sampling
lies within $E_{{\rm bin}}\pm 0.5$,
we adopt this configuration as a ``ground state''.

Thus the ``ground state'' obtained by this procedure always
has an appropriate binding energy.
However, there is still open phase space below the Fermi surface,
since the ``ground state'' is not the energy minimum state of the
Fermions.
In fact, during the time evolution of the ``ground state''
under the QMD dynamics described in previous sub-section,
only 70~\% of the collisions are blocked by the final state
Pauli blocking in the two-body collision term.
It is allowed at a collision that
one nucleon goes down to the lower
energy state and the other goes up to the positive energy state.
This means that some nucleons could be spontaneously emitted
after some time due to the fluctuation of the configuration.
To avoid this problem, we assume from a technical point of view
that any pair of nucleons originated in the same nucleus
does not collide each other until at least one of them experience
a collision with a nucleon from the other nucleus.
By this assumption the number of the emitted nucleons from the
``ground state'' is reduced to less than  about 1~\% of the nucleons
up to the time 200 fm/$c$.

The density profile of the ``ground state'' obtained here has
high density in the center and
rather wide surface shape.
This is due to the large width of Gaussian $L = 2.0 \; {\rm fm}^2$
used in this paper.
On the other hand, the momentum distribution of the ``ground state''
well reproduces the result of Hartree-Fock calculation.
In Fig.\ \ref{denmom}, we show (a) the density $\rho(r)$
distribution and (b) the momentum distribution $\rho(p)$
for the ``ground state''
of $^{40}{\rm Ca}$ obtained by QMD simulation (solid lines).
The results shown here are averaged quantities over time evolution
up to 200 fm/$c$ and over 100 events.
The error bars in Fig.\ \ref{denmom} denote the fluctuations
in time evolution averaged over 100 events.
Although the fluctuation of the one event is much larger,
this figure shows that the ground state profile is very stable
in time on the average over the events.
In the same figure, we also plot the results of Hartree-Fock
calculation (dot-dashed lines) and the limit of infinite nuclear
matter (dashed lines).
The energy spectra of the emitted particles given in the next
section,  particularly of the subthreshold particle production
\cite{niita1}
are more sensitive to the momentum distribution than the density
profile.
This is the reason why we adopted a parametrization which leads to
a better momentum distribution at the cost of diffuse density
profile.

\subsection{Statistical Decay Model}

At the end of the dynamical stage of the reactions,
the QMD simulation yields many fragments,
which are normally in highly excited states.
One may think that the decay process of the excited fragments
might be described by the QMD dynamics
if we can continue the calculation for enough long time.
However, we do not follow this method but instead we stop
the QMD calculation and switch to the statistical decay model (SDM)
at the end of the dynamical stage.
There are two reasons for this hybrid model.
One is that the time scales of
the dynamical and statistical processes are quite different.
It is not clever or even practically impossible to continue
the reliable QMD calculation more than $10^{-20}$ sec
which is necessary to calculate the decay process.
Another is a more fundamental reason that the Fermi statistics,
which is essential to describe the decay process of the fragments,
cannot be traced correctly in the QMD simulation \cite{ohnishi1}.

We identify the fragments together with
their excitation energies
at about 100 $\sim$ 150 fm/$c$ of the QMD simulation.
The dependence of the final results on this switching time
will be discussed in the next section.
Each fragment is recognized by using a minimum distance chain
procedure, i.e.,
two nucleons are considered to be bound in a fragment if the
distance
between their centroids is smaller than 4 fm.
We then calculate the total energy of the fragment in its rest frame
and estimate the excitation energy by subtracting
the ground state energy given by Eq.\ (\ref{liquid-e}).

Though there have been proposed many sophisticated statistical
decay codes so far,
we use here the simple model of light particle evaporation.
We consider only $n$, $p$, $d$, $t$, $^3$He, and $\alpha$
evaporation.
The emission probability $P_x$ of these particles $x$ is given with
the Fermi gas model as,
\begin{equation}
P_x = (2J_x +1) \,
m_x \, \epsilon  \, \sigma_x(\epsilon) \, \rho(E) \, d\epsilon ,
\label{prob1}
\end{equation}
where $J_x, m_x,$ and $\epsilon$ are the spin, mass and kinetic
energy of the particle $x$, while $\sigma_x(\epsilon)$ and
$\rho(E)$ denote the inverse cross section
for the absorption of the particle with energy $\epsilon$
and the level density of the residual
nucleus with the excitation energy $E$, respectively.
We use the following simple form for $\rho(E)$,
\begin{equation}
\rho(E) = w_0 \, \exp\left( 2 \sqrt{a\, E} \right),
\label{level1}
\end{equation}
with $a = A/8 \; {\rm MeV}^{-1}$ and $w_0$ is a constant.
The inverse cross section is assumed to have the form
\begin{equation}
\sigma_x(\epsilon) = \left\{
\begin{array}{lr}
\left(1-U_x/\epsilon\right) \, \pi R^2 &
: \epsilon > U_x \\ [1ex]
\; 0 & : \epsilon \le U_x \\
\end{array}
\right.
\label{invers}
\end{equation}
where $R$ denotes the absorption radius and
$U_x$ is a Coulomb barrier for the particle $x$,
for which we employ empirical values used in the existing
statistical decay code \cite{gemini}.
The excitation energy $E$ in Eq.\ (\ref{level1}) is given by
\begin{equation}
E = E_0 - \epsilon - Q,
\label{excite}
\end{equation}
where $E_0$ denotes the excitation energy of the parent nucleus
and $Q$ is the reaction Q-value calculated
from the mass formula Eq.\ (\ref{liquid-e}).
The total emission probability $R_x$ of the particle $x$ is obtained
by integrating the available energy of Eq.\ (\ref{prob1}) as
\begin{eqnarray}
R_x & = & (2J_x+1)  \, m_x \, \nonumber \\[1ex]
& & \times \int_{U_x}^{E_0-Q_x} \,
\epsilon \, \sigma_x(\epsilon) \, \rho(E_0-Q_x-\epsilon)
\, d\epsilon.
\label{totalp}
\end{eqnarray}
This integration can be calculated analytically and the
energy spectrum of the emitted particles is given by
\begin{equation}
N(\epsilon_x) \, d\epsilon_x  =  \frac{\epsilon_x - U_x}{T_x^2}
\, \, \exp\left\{ - \frac{\epsilon_x - U_x}{T_x}
\right\} \, \, d\epsilon_x,
\label{spec1}
\end{equation}
with
\begin{equation}
a\, T_x^2  =   E_0 - U_x - Q_x.
\label{spec2}
\end{equation}

In this formulation, we do not consider the $\gamma$ decay nor
the angular momentum dependence.
The latter is important for the heavy-ion reactions but
not so serious for the nucleon-induced reactions considered
in this paper.
We simulate the whole statistical decay process as a sequential
light particle evaporation discussed above by making use of the
Monte-Carlo method until no more particle can be emitted.

\section{Analysis of the (N,xN') Reactions}

In this section, we systematically apply the QMD plus SDM method
described in the previous section to $(N,xN')$
(nucleon in, nucleon out) type reactions.
In order to get sufficient statistics, we performed the QMD
calculations for a large number of events, typically 50000 events,
and averaged them to obtain the following results.

We first check the dependence on the switching time $t_{{\rm sw}}$
when the QMD calculation is stopped and switched to
SDM, which is an ambiguous point of the present model.
In Fig.\ \ref{pb15001}, we show a typical neutron energy
spectrum at $30^{\circ}$ laboratory angle for the reaction
$p(1.5 {\rm GeV}) + ^{208}$Pb.
Note that x-axis is plotted in a logarithmic scale to compare
the calculated results in detail with the experimental data
particularly
in a low energy region.
The experimental data (full boxes with error bars) are taken from
Ref.\ \cite{ishibashi}.
The solid histogram denotes the final result of the QMD + SDM
calculation.
In this case,  we switch the QMD calculation to SDM at 100 fm/$c$.
In the same figure, we also plot the spectrum of the neutron
obtained only by the QMD calculation up to 100 fm/$c$
(dot-dashed histogram)
and that coming from the QMD fragments calculated with SDM
(dashed histogram), respectively.
The former shows a cascade and/or preequilibrium energy spectrum,
while the latter an evaporation spectrum.
These two components of spectrum are affected by changing
the switching time $t_{{\rm sw}}$.
However, the total spectrum shape,
which is the sum of the two components,
stays almost unchanged
if an adequately long time is chosen for the switching time
$t_{{\rm sw}}$.
This is shown in Fig.\ \ref{pb15002}, where we plot results of
the total spectra calculated by QMD + SDM
with three different switching times, 50 fm/$c$ (dashed line),
100 fm/$c$ (solid line) and 150 fm/$c$ (dot-dashed line).
This figure shows that although the latter two lines resemble
each other, they deviate definitely from the first line.
This indicates that the QMD fragments before 100 fm/$c$ are not in
thermal equilibrium
and that within a time interval from 100 fm/$c$ to 150 fm/$c$
the decay processes of the excited fragments described by QMD
and SDM are nearly equivalent.
Although we should keep in mind that these two are not identical
at low temperature as the former is always dominated by the
classical statistics \cite{ohnishi1},
we can conclude that
the final results are not sensitive to the switching
time $t_{{\rm sw}}$ as long as it is chosen
after the time when the thermal equilibrium is achieved
and before the time the temperature of the fragments
become low and classical statistics breaks down seriously.
The similar conclusion has been obtained in Ref.\ \cite{maru92a},
which indicates that the minimum switching time
to get stable results depends on the size of system and the
incident energy.
For the case of nucleon induced reactions, we found that
100 fm/$c$ is enough to get stable results and we use this value
for all systems in the present study.

Next check is the detailed examination of the inelastic channels
in the two-body collision term.
For this purpose we compare our results with the experimental
data at high incident energy of proton on the light-mass target,
which directly reflects the elementary processes included
in the model.
In Fig.\ \ref{enyo1}, we plot the invariant cross section of
the proton (left-hand-side) and negative pion (right-hand-side)
emission for the reaction $p \,(3.17 {\rm GeV}) +^{27}$Al.
The experimental data (full boxes with error bars) are taken
from Ref.\ \cite{enyos} and the results of QMD + SDM are
denoted by solid histograms.
In the same figure, we plot results of QMD + SDM
with the different choice of the
angular distribution of the resonances.
The dashed histograms are the results obtained with only
the resonance-like angular distribution of Eq.\ (\ref{angwolf1}),
while the dot-dashed histograms are those with the direct-like
angular distribution of Eq.\ (\ref{angdir1}), respectively.
This figure shows that the average of the two components
of the angular distribution of
Eq.\ (\ref{angtot}) well fits
the experimental proton spectra.
On the other hand, the pion spectra are rather insensitive
to the angular distribution of the resonances.
Instead, their spectra are dominantly determined by the mass
distribution of the resonances of Eq.\ (\ref{masdis1}).
Although the authors of Ref.\ \cite{enyos} analyzed these data
by making use of the two-moving-source model,
our QMD + SDM can reproduce excellently the proton and pion
spectra at the same time without any special assumption.

In Figs.\ \ref{fe113}-\ref{pb3000}, we compare the neutron energy
spectra obtained by the QMD + SDM calculations
with the experimental data \cite{ishibashi,Meier}
for Fe and Pb targets at the proton energy from
113 MeV up to 3 GeV.
In the fields of application of accelerators,
such as spallation neutron sources, accelerator-based
transmutation systems,
and shielding
of cosmic ray in space activity,
the production of slow neutrons plays an important role.
That is the reason we chose these data \cite{ishibashi,Meier}
to compare with, since the neutron spectra
from 1 MeV up to the beam energy are available.
For efficient comparisons of the calculations and the data,
both for low energy and high energy regions,
we plot the same results in two figures with the x-axis in
a logarithmic scale (left-hand-side)
and in a linear scale (right-hand-side).
One can see the detail of
the thermal and preequilibrium neutron spectra
in the former figure,
while the direct-like components of the spectra in the latter
figure, respectively.

These figures indicate that over the broad range of
the incident energies from 100 MeV to 3 GeV,
independently of the targets, and of all angles of the outgoing
neutrons, our results of the neutron energy spectra
agree well with the data from 1 MeV up to the beam energy.
Though one may notice some disagreement at the high energy part of
the most forward angle, which will be discussed later,
the overall agreement is satisfactory.
Particularly a remarkable agreement of the present calculations
with the low energy neutron data below several tens MeV confirms
that the QMD gives  proper excitation spectra of the excited
fragments from which the statistical neutron emission takes place.
With a suitable chosen fixed set of parameters, our QMD plus SDM
model is able to reproduce quantitatively the overall neutron spectra
for the broad range of the incident energies and target masses.

At 113 MeV, we additionally compare our results with the prediction
of the intranuclear cascade plus light particle evaporation model
(NUCLEUS) \cite{nucleus}.
This model is essentially the intranuclear reaction part of
NMTC \cite{nmtc} and HETC \cite{hetc} codes.
Calculations with NUCLEUS yield almost the same results as ours
in the forward angles but give lower values in the backward angle,
which is denoted by the dashed histogram at 150 degree
in Fig.\ \ref{fe113}.
In this energy regime it has been reported \cite{blann1,blann2}
that the semiclassical preequilibrium models based on
an intranuclear cascade model also fail to reproduce the angular
distributions.
We found from the detailed comparison of the calculations
that the underestimation of the backward angle in the above
models is due to the insufficient treatment of
the soft interaction of a nucleon with all the rest of the nucleons
in the nucleus,
which is naturally included in the QMD formalism.

One may think that this explanation is
in contradiction with
Ref.\ \cite{peilert,blann3} where
they attribute the failure to the insufficient contributions
from second and higher order collisions.
To resolve this problem,
we have checked that the NUCLEUS has almost the same prescription of
hard nucleon-nucleon interaction
and has almost the same momentum distribution in the ground state
as QMD.
Difference is that our QMD part includes the soft nucleon-nucleon
interaction but NUCLEUS does not.
This soft interaction diffracts the nucleon.
As a result, the yields of the backward angle increase
and by the same reason
the number of multiple hard collisions also increases.

On the other hand,
a multistep model of
the Feshbach-Kerman-Koonin (FKK) \cite{fkk}
has been also applied to the energy regime
around 100 MeV \cite{blann3}.
Although the FKK approach successively reproduced
the angular distribution,
the overall absolute values of their results
are very sensitive to the strength parameter of
the residual interaction,
which is adjusted to fit the experimental data.
The strength so determined depends on the incident particle,
target nucleus and incident energy.
In QMD, on the contrary, the parameters of the soft
nucleon-nucleon interaction in Eq.~(\ref{ham1})
are taken common to all reactions and
determined from the nuclear saturation condition.
In addition,
the final results are not so sensitive to them.

The first analysis of $(p,xn)$ reactions by the QMD approach
in the energy regime up to 800 MeV
has been done by Peilert et al. \cite{peilert}.
The neutron spectra of their results are very similar to
those of the present work above several tens MeV.
Their analysis, however, cannot predict a whole spectra of
neutron, since the contribution of the statistical decay
from the excited fragments produced in the QMD calculation
was not considered in their work.

Though the present results show overall agreement with the data
for the very broad energy regime,
one can see a systematic deviation from the data in the high
energy part of the neutron spectra at the most forward angle
at incident energy from 113 MeV up to 800 MeV
(see the right-hand-side of Fig.\ \ref{fe113}-\ref{pb800}).
We suppose that the soft nucleon-nucleon interaction
is responsible for this deviation.
One possibility is a momentum dependent interaction
that is not included in the present QMD,
by which
the nucleon could be affected coherently by the surrounding
nucleons when its momentum is drastically changed
by the hard nucleon-nucleon scattering.
For the higher incident energies (see Fig. \ref{pb1500} and
\ref{pb3000}), this deviation disappears.
In those cases, we have checked that the neutrons
in the high energy part of the forward angle emerge after
at least once experiencing the resonances of nucleon, and that
the effects of the soft interaction is relatively small.
The analysis by the QMD including the momentum dependent interaction
will be reported in the forthcoming paper.

\section{Summary and Conclusion}

We have proposed the quantum molecular dynamics (QMD) incorporated
with the statistical decay model (SDM)
aiming to describe various nuclear
reactions in a unified way,
and applied this model to the $(N,xN')$ reactions.
We have checked and found that the final results do not depend on
the switching time when the QMD simulation is stopped and switched
to the SDM calculation as long as the switching time is chosen
between 100 fm/$c$ and 150 fm/$c$ for the nucleon induced reactions.
Therefore there left little ambiguity
with respect to the switching of two different
kinds of models to describe whole process of the reactions
in a unified way.

In order to describe the reactions at high incident energies up to
3 GeV,
we have taken into account two baryonic resonances,
the $\Delta$(1232)
and the $N^*$(1440) as well as the pions in the QMD model.
The elementary cross sections related to these resonances and pions
are basically taken from the experimental data.
The angular distributions of the resonances, which information is
very poor in the experimental data, have been fixed to fit the
$^{27}$Al$(p,p')$ data \cite{enyos} at 3.17 GeV.
It should be noted that the energy spectra of the nucleons from the
$(N,xN')$ reactions on the small target are suitable quantities
to obtain the detailed information of the angular distribution
of the resonances,
while the pion spectra are useful to extract the information of
the mass distribution of the resonances.

In addition to the relativistic kinematics and approximately
covariant prescription of the collision term,
we have introduced the Lorentz-scalar quantities to the arguments
of the interactions and to the phase space factor.
By this relativistic corrections together with the Lorentz
contraction of the
initial phase space distribution, the main part of the relativistic
dynamical effects is approximately described in our QMD
for the energy regime up to 3 GeV/nucleon.
Validity of this model has been confirmed by the analysis
of the transverse flow for the
heavy-ion collisions in comparison with the results obtained by
the covariant version of quantum molecular dynamics (RQMD).

We have applied systematically QMD + SDM to the $(N,xN')$ reactions
for a broad range of incident energies from 100 MeV to 3 GeV and
of target masses.
The present model reproduced the overall
feature of the outgoing neutrons quite well
without assuming any reaction mechanism,
and without changing a parameter set.
Although there are a lot of parameters in the model
which have not been investigated extensively in this paper,
the final neutron spectra analyzed here do not depend so much
on them,
for example, the equation of state (choice of the interaction),
width of the gaussian wave packets, and the detail of the
statistical decay process.
The main ingredients of the model,
which produce the present results of the neutron
spectra down to the energy of several MeV,
are the parametrization of the elastic and inelastic
elementary cross section
and the many-body dynamics itself,
which have been discussed both in detail in this paper.
We thus conclude that the present QMD + SDM scheme gives
a unified picture
of the major three reaction mechanisms of $(N,xN')$ reactions;
i.e. compound, pre-equilibrium and spallation processes.

Finally, we should mention that the present model is ready to be
applied directly to the heavy-ion reactions in its original form.
A study of heavy-ion reaction using this model is now under
consideration.

\acknowledgements

The authors are grateful to  Profs. K. Ishibashi and M. M. Meier for
supplying us with the experimental data prior to the publication.


%
%
%
\begin{figure}
\caption{$p$-$p$ (a) and $p$-$n$ (b) total (solid line),
         elastic (long dashed line),
         and inelastic (dot-dashed line) cross sections,
         which is the sum of the $\Delta$ (short dashed line)
         and $N^*$
         (dotted line) production cross section, calculated by
         Eqs.\ (\protect\ref{signn1},
         \protect\ref{signn2},
         \protect\ref{wests},
         \protect\ref{delta},
         \protect\ref{nstar}).
         The experimental total (open circles), elastic
         (open triangles) and inelastic (open boxes) cross sections
         are taken from Ref.\ \protect\cite{data1}.
         }
\label{pptot}
\end{figure}
\begin{figure}
\caption{Pion cross sections of
         $pn \to nn\pi^+ + pp\pi^-$ (solid line).
         The gray bold lines denote the results of the original
         parametrization of VerWest and Arndt \protect\cite{west},
         while the experimental data are taken from
         Ref.\ \protect\cite{data1}.
        }
\label{pppi}
\end{figure}
\begin{figure}
\caption{Two pion production cross section.
         The solid line denotes $\sigma_{N^*}$,
         while the experimental data \protect\cite{data1}
         are scaled according to the factors of
         Eq.\ (\protect\ref{twopi}).
        }
\label{signn3}
\end{figure}
\begin{figure}
\caption{The energy dependent angular distribution of
        resonance production.
        (a) Two components of the angular distribution.
        The solid lines represent the elastic like angular
        distributions of Eq.\ (\protect\ref{angdir1}),
        $ \frac{1}{2} [ g_{\rm D}(s,\cos\theta)
        + g_{\rm D}(s,-\cos\theta)] $,
        while the gray bold dashed lines are the resonance like
        angular distributions of Eq.\ (\protect\ref{angwolf1}).
        (b) The total angular distribution of $\Delta$ (solid lines),
        which is the sum of $g_{\rm R}$ and $g_{\rm D}$.
        The gray bold dashed lines are the same as in (a)
        }
\label{angf}
\end{figure}
\begin{figure}
\caption{The directed transverse momentum as a function of
        energy/nucleon
        for $^{40}{\rm Ca}+^{40}{\rm Ca}$ reaction
        at $b$ = 2~fm.
        The results are shown as the differences from that of
        the RQMD \protect\cite{rqmdt}.
        The solid line with full boxes
        denotes the result of the present model with
        the relativistic corrections and the initial Lorentz
        contraction.
        The dot-dashed line with open circles is the result of
        the standard QMD without the initial Lorentz contraction
        and without the relativistic corrections,
        while the dashed line with open boxes is
        the result of the standard QMD only
        with the initial Lorentz contraction,
        respectively.
        }
\label{fig-flow}
\end{figure}
\begin{figure}
\caption{(a) Density distribution $\rho(r)$
        and (b) momentum distribution $\rho(p)$
        of the ground state of $^{40}{\rm Ca}$ obtained by the
        QMD simulation (solid lines) averaged over the time
        evolution
        up to 200 fm/$c$ and over 100 events.
        The error bars  denote the fluctuations
        in time evolution averaged over 100 events.
        The dot-dashed lines show the results of Hartree-Fock
        calculations, while the dashed lines denote the limit of
        infinite nuclear matter.
        }
\label{denmom}
\end{figure}
\begin{figure}
\caption{Neutron energy spectrum at the $30^{\circ}$ laboratory
        angle for the reaction $p(1.5 {\rm GeV}) + ^{208}$Pb.
        The experimental data (full boxes with error bars) are
        taken from Ref.\ \protect\cite{ishibashi}.
        The solid line denotes the final result of
        the QMD + SDM calculation with the switching time
        $t_{{\rm sw}} = 100$ fm/$c$.
        The dot-dashed line is the result obtained only by the QMD
        calculation up to 100 fm/$c$, while the dashed
        line the neutron spectra  coming from the QMD
        fragments calculated with SDM, respectively.
        }
\label{pb15001}
\end{figure}
\begin{figure}
\caption{The total energy spectra calculated by QMD + SDM
        with different switching time,
        $t_{{\rm sw}} = 50$ fm/$c$ (dashed line),
        $t_{{\rm sw}} = 100$ fm/$c$ (solid line) and
        $t_{{\rm sw}} = 150$ fm/$c$ (dot-dashed line).
        The reaction system is the same as
        in Fig.\ \protect\ref{pb15001}.
        }
\label{pb15002}
\end{figure}
\begin{figure}
\caption{Invariant cross sections of the proton (left-hand-side)
        and negative pion (light-hand-side) emission for the
        reaction
        $p(3.17 {\rm GeV}) +^{27}$Al at different laboratory angles
        as indicated in the figure. Full boxes with error bars
        are the experimental data taken from
        Ref.\ \protect\cite{enyos}
        and the results of QMD + SDM are denoted by solid
        histograms.
        The dashed histograms are results of another QMD + SDM
        obtained with only the resonance-like angular distribution
        Eq.\ (\protect\ref{angwolf1}), while the dot-dashed
        histograms
        with the direct-like angular distribution
        Eq.\ (\protect\ref{angdir1}), respectively.
        }
\label{enyo1}
\end{figure}
\begin{figure}
\caption{Neutron energy spectra
        for the reaction $p\,$(113 MeV)$+^{56}$Fe at
        different laboratory angles as indicated in the figure.
        The x-axis is plotted in a logarithmic scale in the
        left-hand-side, while the linear scale in the
        right-hand-side.
        The solid histograms are the results of QMD + SDM,
        the open circles
        with error bars denotes the experimental data taken from
        Ref.\ \protect\cite{Meier}. The dashed histograms
        denote the results of
        NUCLEUS \protect\cite{nucleus} at the $150^{\circ}$
        laboratory angle.
        }
\label{fe113}
\end{figure}
\begin{figure}
\caption{Same as Fig.\ \protect\ref{fe113}
        for the reaction $p\,$(256 MeV)$+^{208}$Pb.
        }
\label{pb256}
\end{figure}
\begin{figure}
\caption{Same as Fig.\ \protect\ref{fe113}
        for the reaction $p\,$(597 MeV)$+^{56}$Fe.
        }
\label{fe597}
\end{figure}
\begin{figure}
\caption{Same as Fig.\ \protect\ref{fe113}
        for the reaction $p\,$(800 MeV)$+^{208}$Pb.
        }
\label{pb800}
\end{figure}
\begin{figure}
\caption{Same as Fig.\ \protect\ref{fe113}
        for the reaction $p\,$(1500 MeV)$+^{208}$Pb.
        The full boxes with error bars are the experimental data
        taken from Ref.\ \protect\cite{ishibashi}.
        }
\label{pb1500}
\end{figure}
\begin{figure}
\caption{Same as Fig.\ \protect\ref{pb1500}
        for the reaction $p\,$(3000 MeV)+$^{208}$Pb.
        }
\label{pb3000}
\end{figure}
%
%
%
\begin{table}
\caption{Elastic cross section parameters.}
\begin{tabular}{ccc}
  & $p$-$n$ & others \\
\tableline
$C_1$ (mb) & 28.0 & 35.0 \\
$C_2$ (mb) & 27.0 & 20.0 \\
$C_3$ (mb) & 12.34 & 9.65 \\
$C_4$ (mb) & 10.0 & 7.0 \\
\end{tabular}
\label{table1}
\end{table}
\begin{table}
\caption{Inelastic cross section parameters.}
\begin{tabular}{cccc}
  & $\sigma_{11}$ & $\sigma_{10}$ & $\sigma_{{\rm N}^*}$ \\
\tableline
$\alpha$    & 3.0 & 14.0 & 23.0 \\
$\beta$     & 0.9 & -0.3 & 1.5  \\
$m_0$ (MeV) & 1188 & 1245 & 1472 \\
$\Gamma$ (MeV) & 99.02 & 120.0 & 300.0 \\
\end{tabular}
\label{table3}
\end{table}
\begin{table}
\caption{Parameters in the width of resonances.}
\begin{tabular}{lccc}
  & $M_r$ (MeV) & $\Gamma_r$ (MeV) & $\beta_r$ (MeV) \\
\tableline
$\Delta$    & 1232 & 110 & 300 \\
$N^*$       & 1440 & 200 & 523 \\
\end{tabular}
\label{table4}
\end{table}
\begin{table}
\caption{Parameters in $f_R(s,\cos\theta)$.}
\begin{tabular}{lccc}
$\sqrt{s}$(GeV)  & $\sqrt{s}\le 2.14$  & $2.14 < \sqrt{s}
\le 2.4$ & $2.4 \le \sqrt{s}$ \\
\tableline
$a_1(s)$    & 0.5 & $29.03-23.75s+4.865s^2$  & 0.06 \\
$a_3(s)$    & 0.0 & $-30.33+25.53s-5.301s^2$ & 0.4 \\
\end{tabular}
\label{table5}
\end{table}
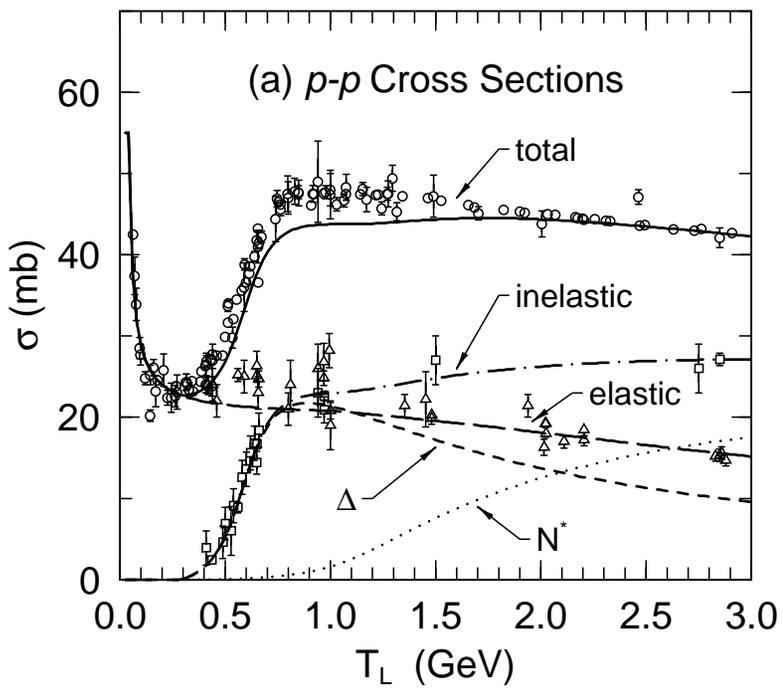
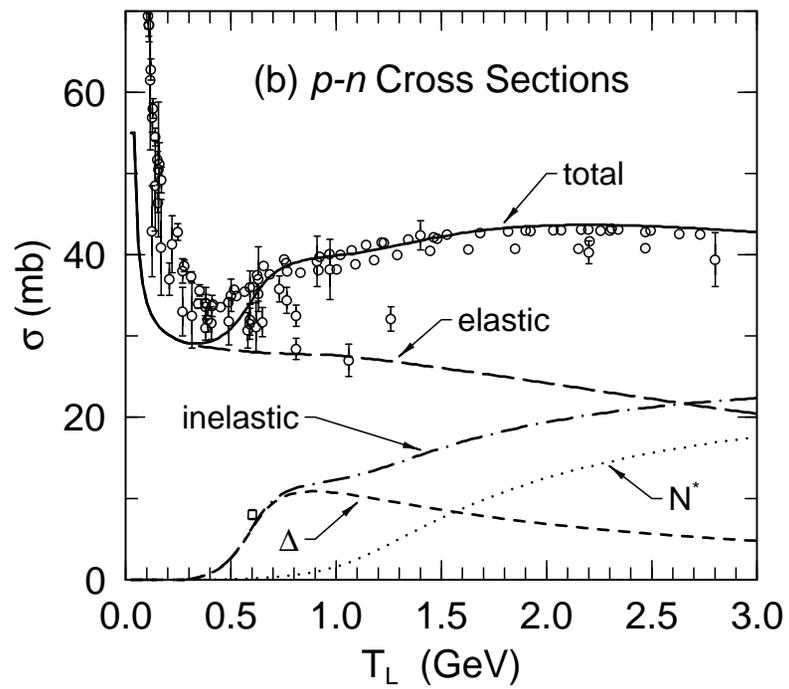
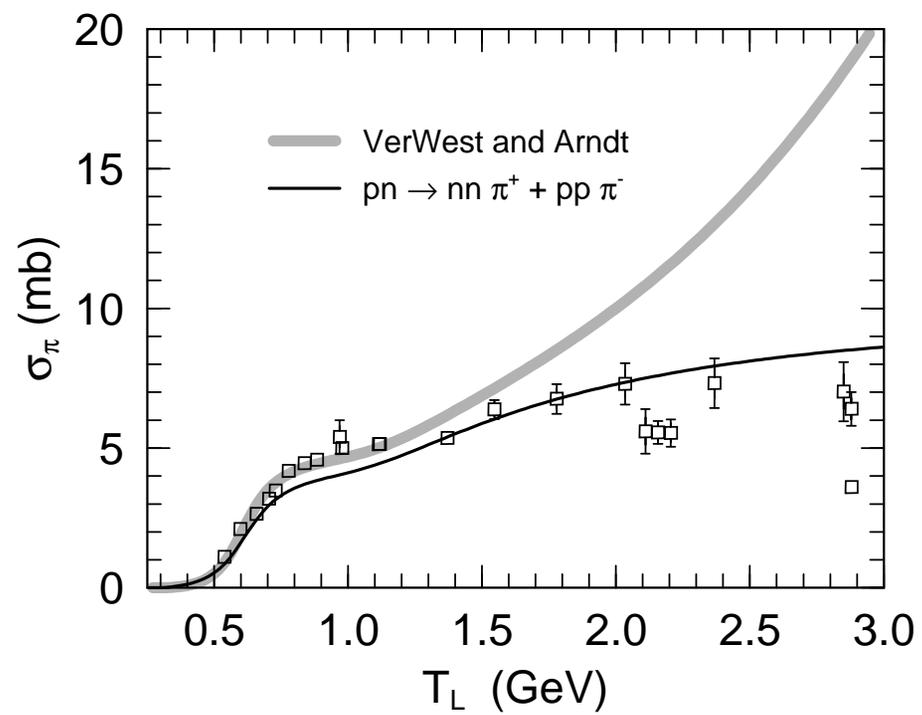
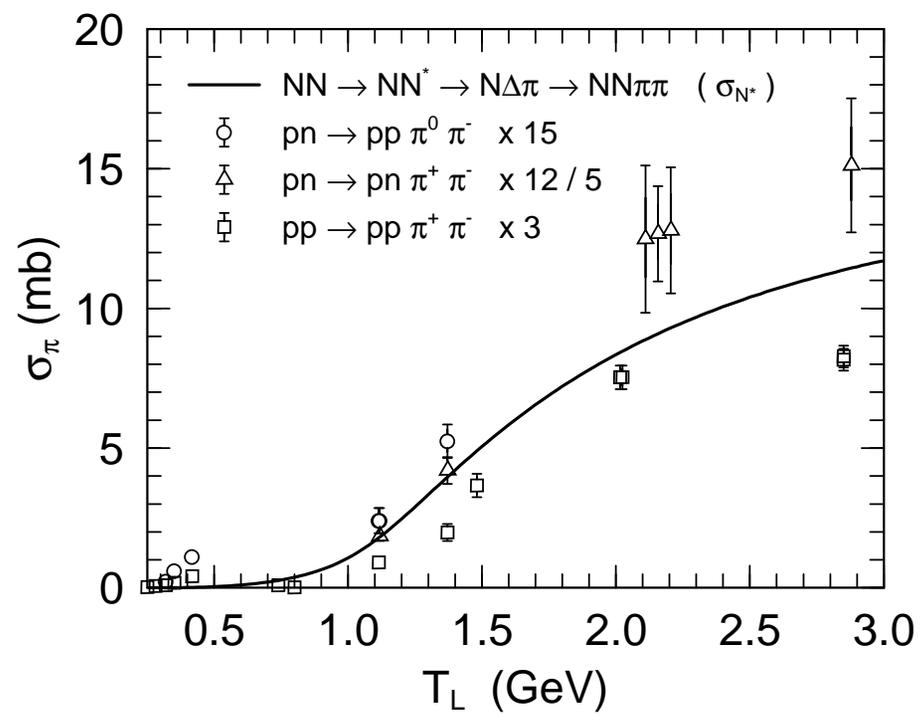
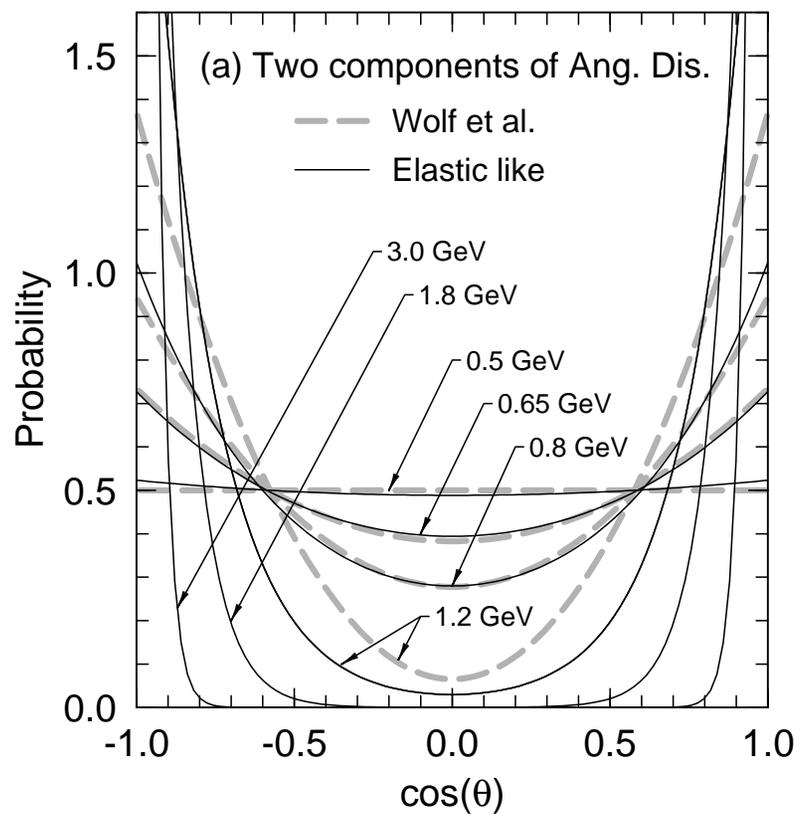
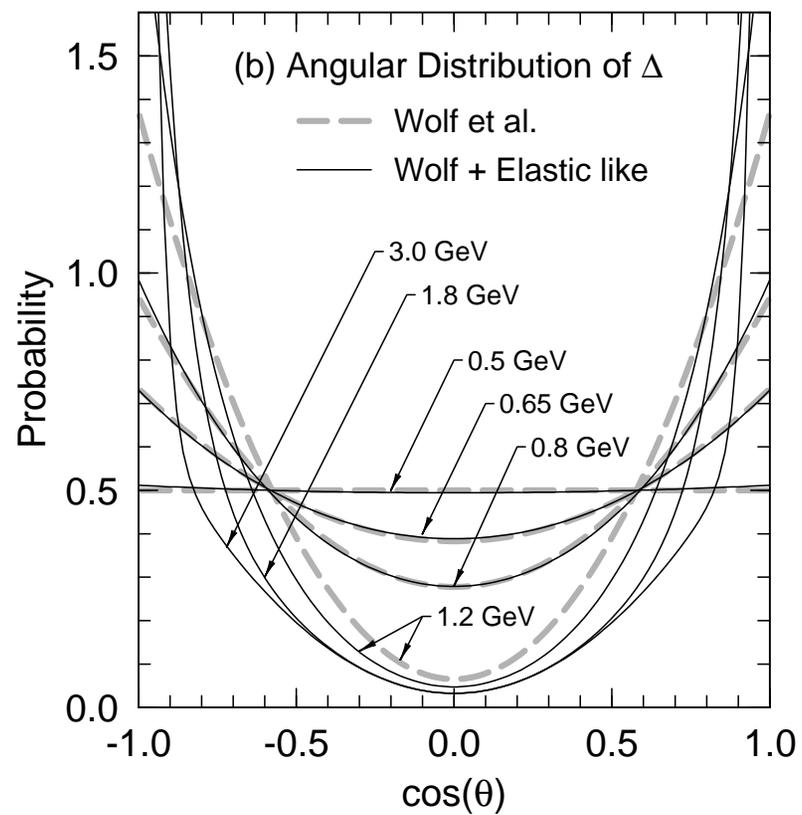
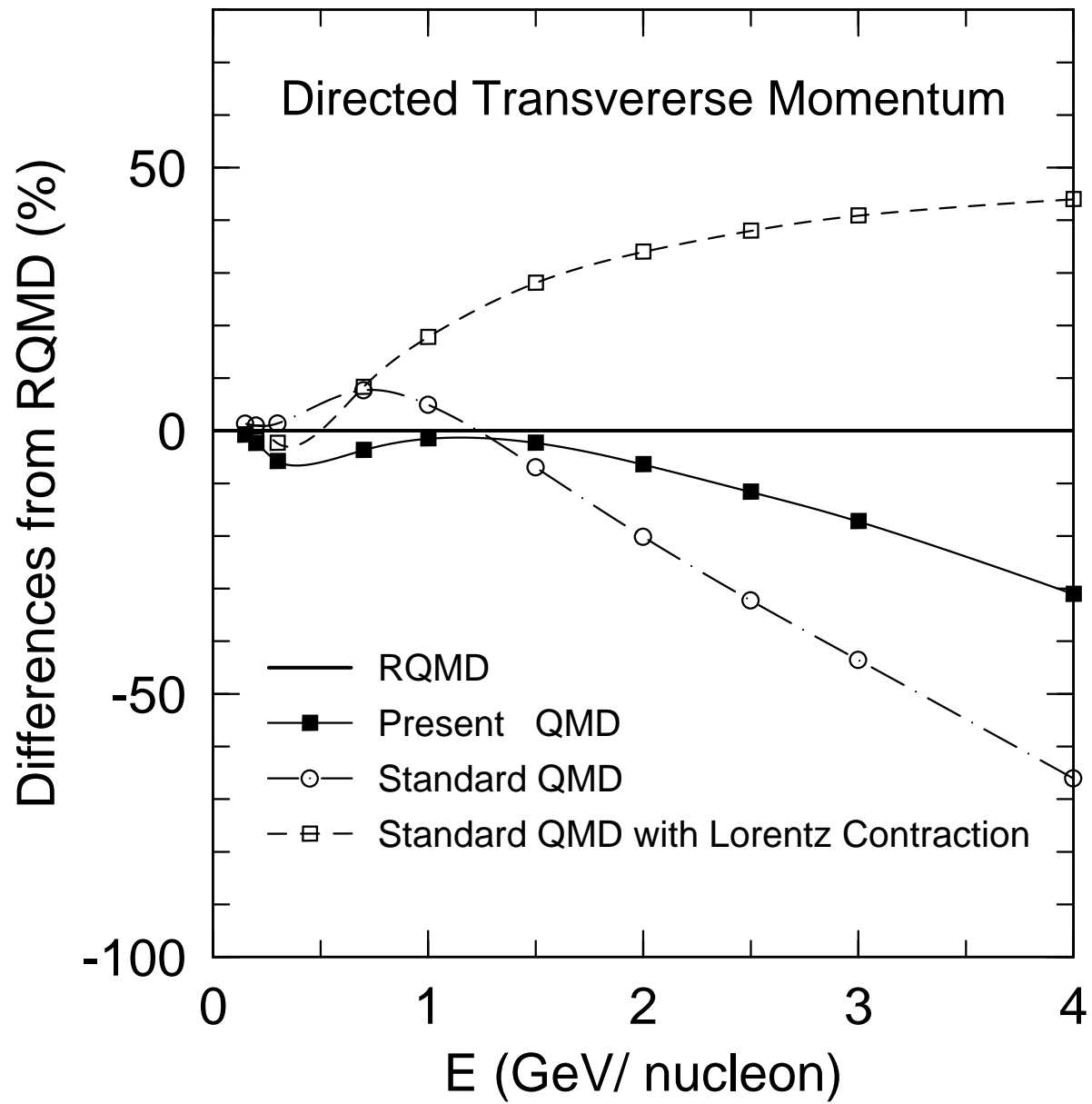
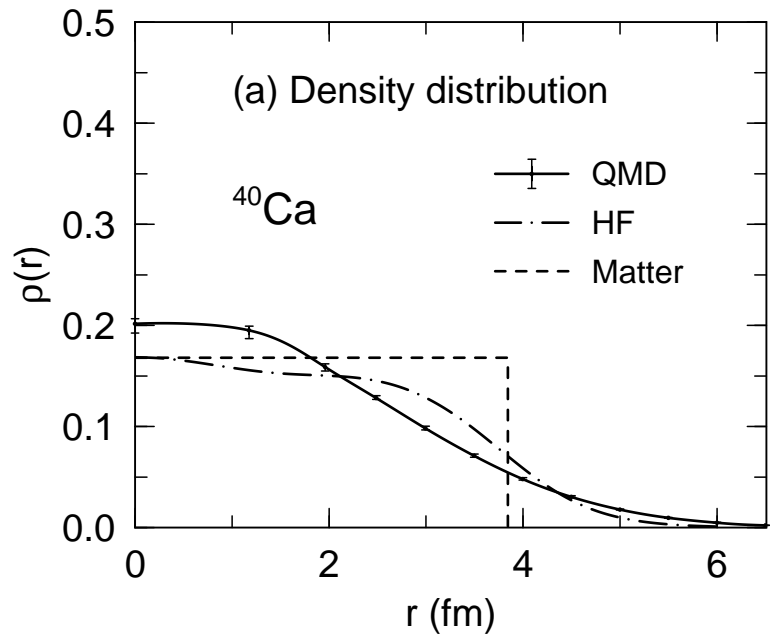
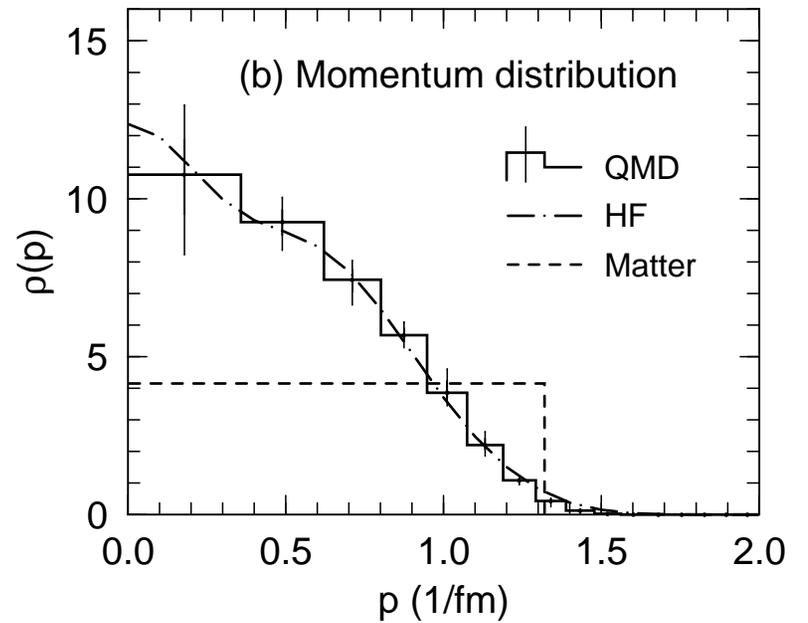
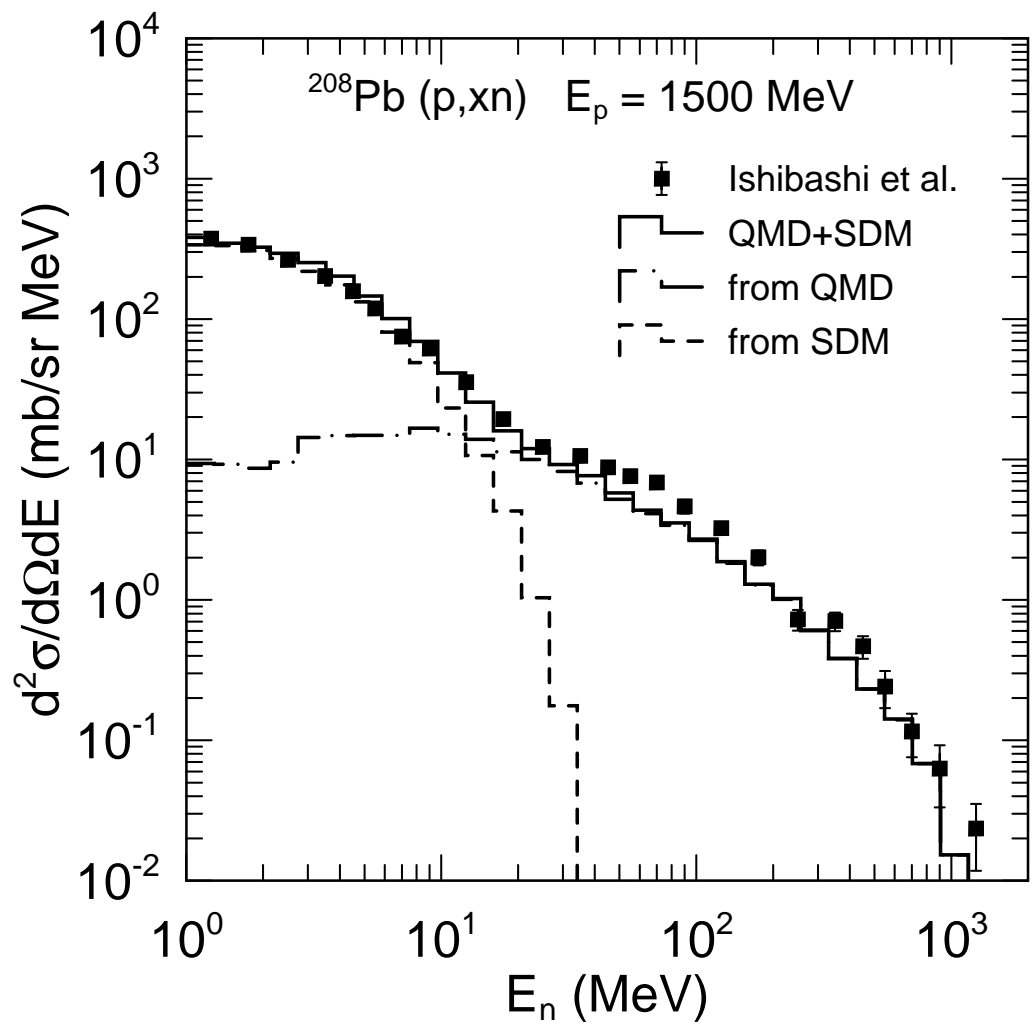
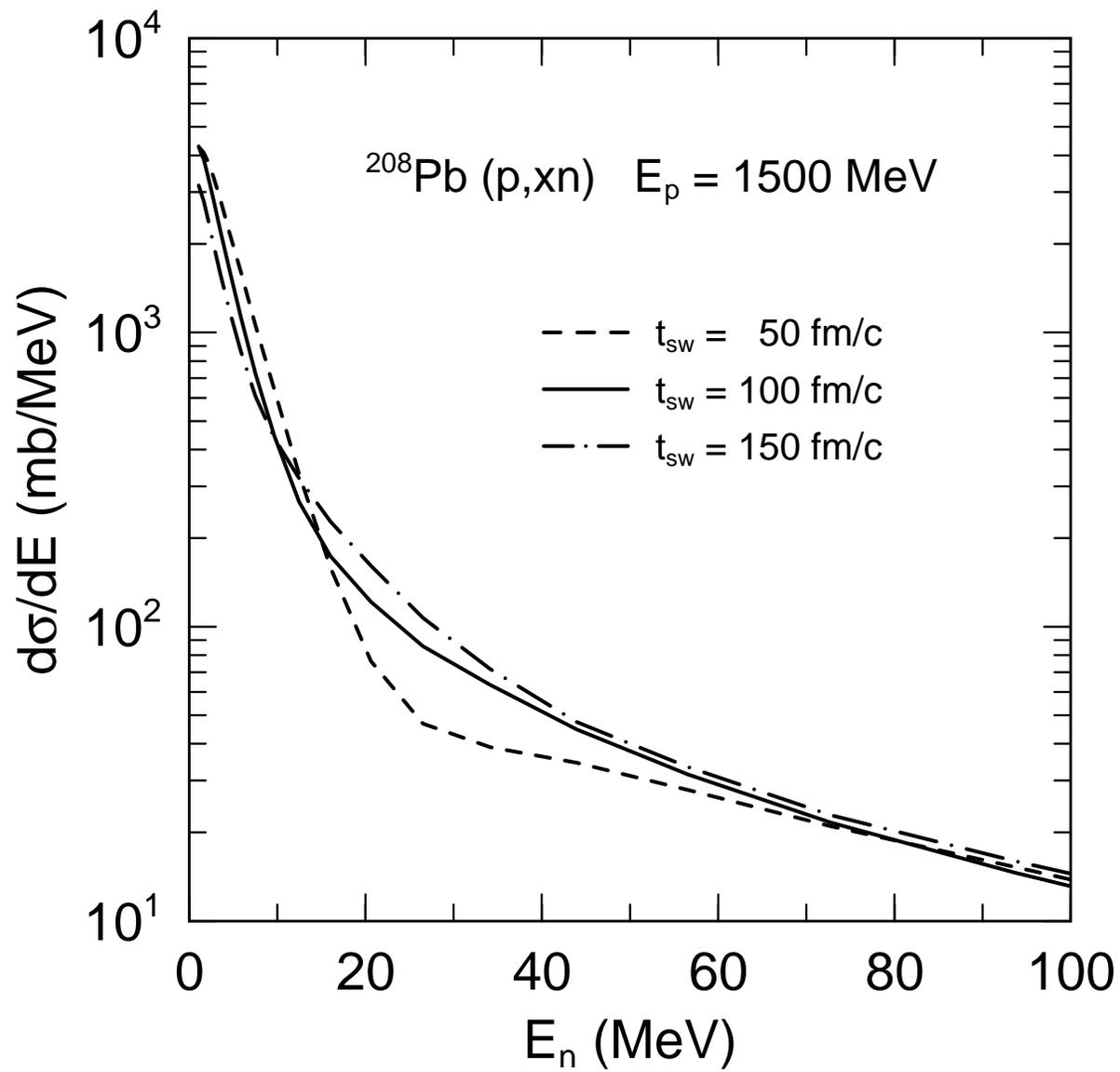
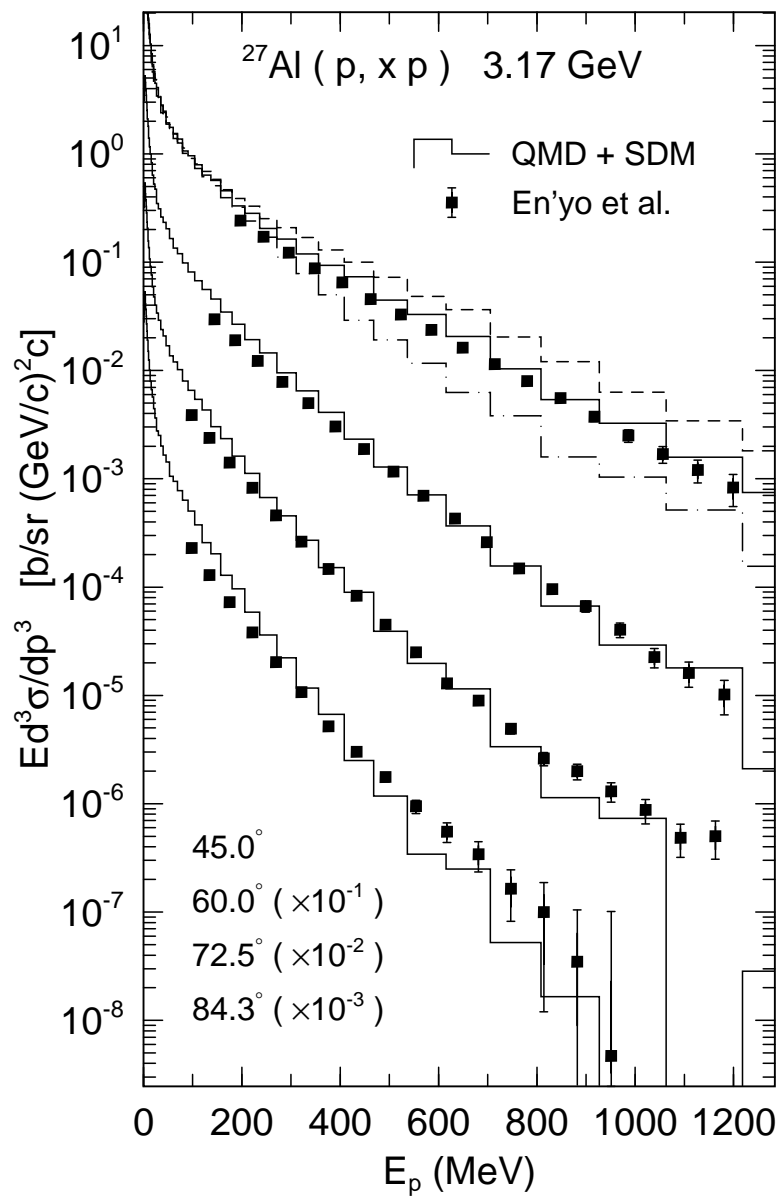
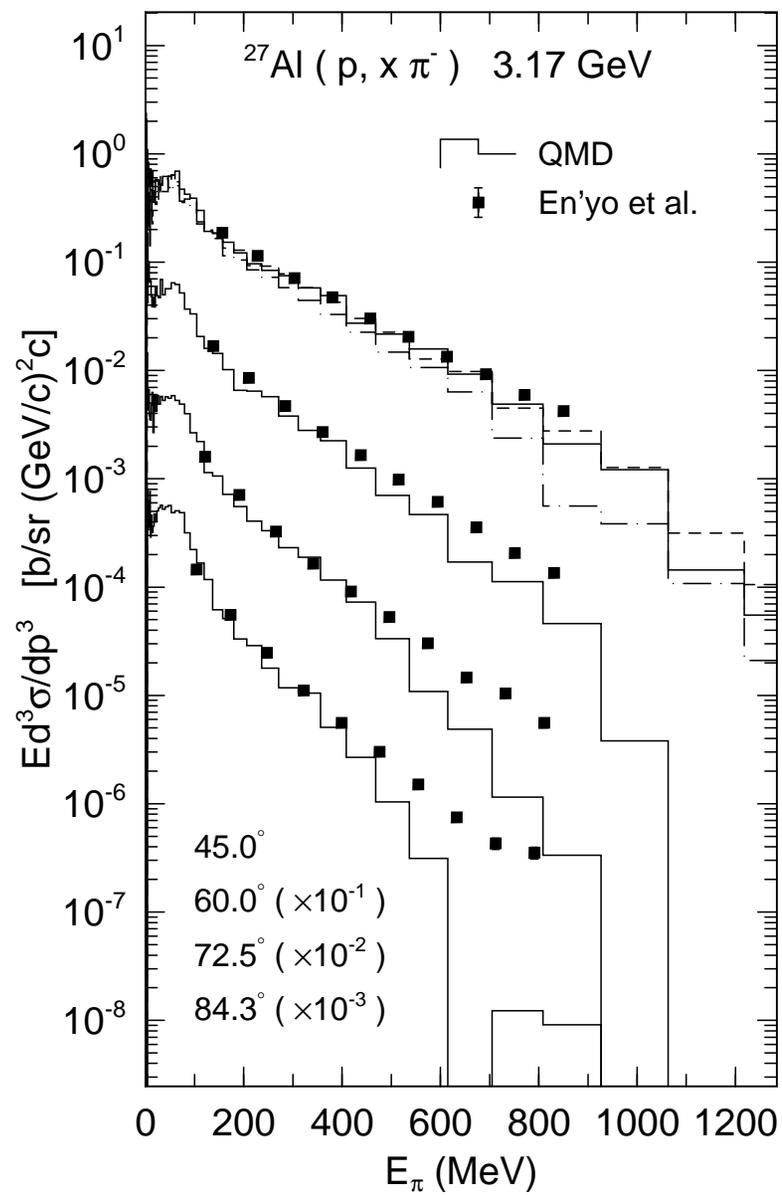
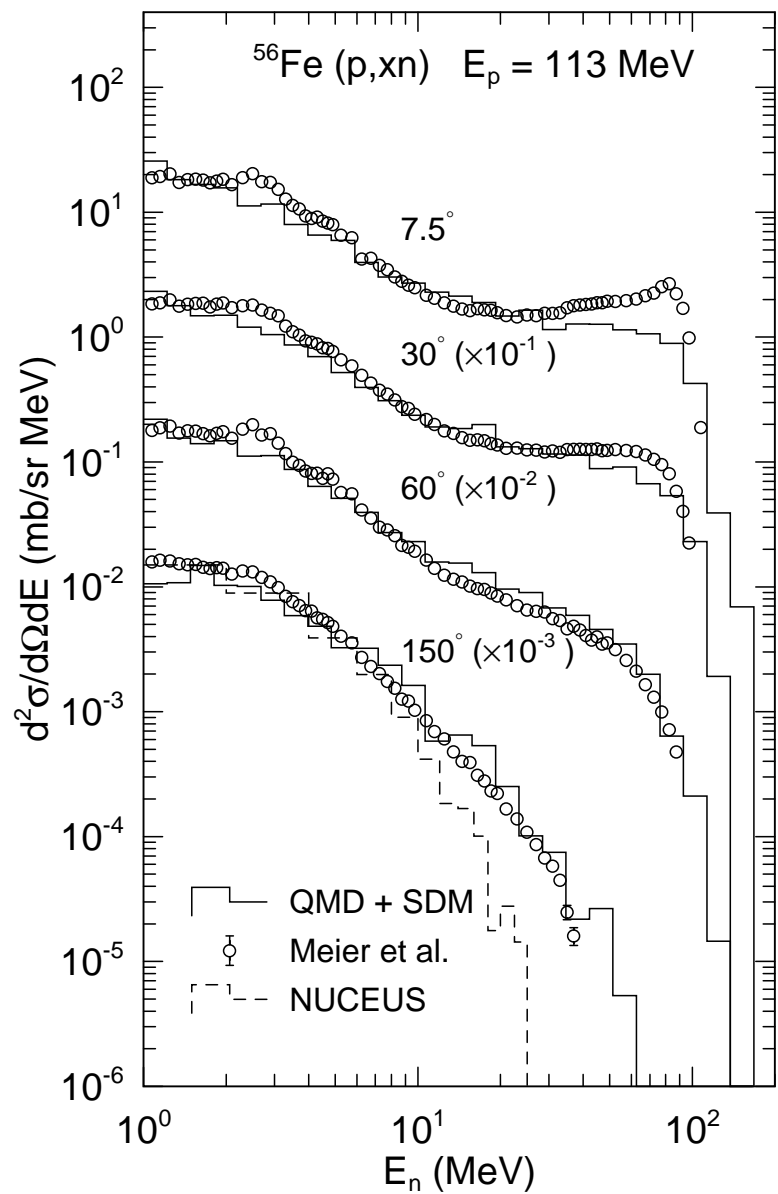
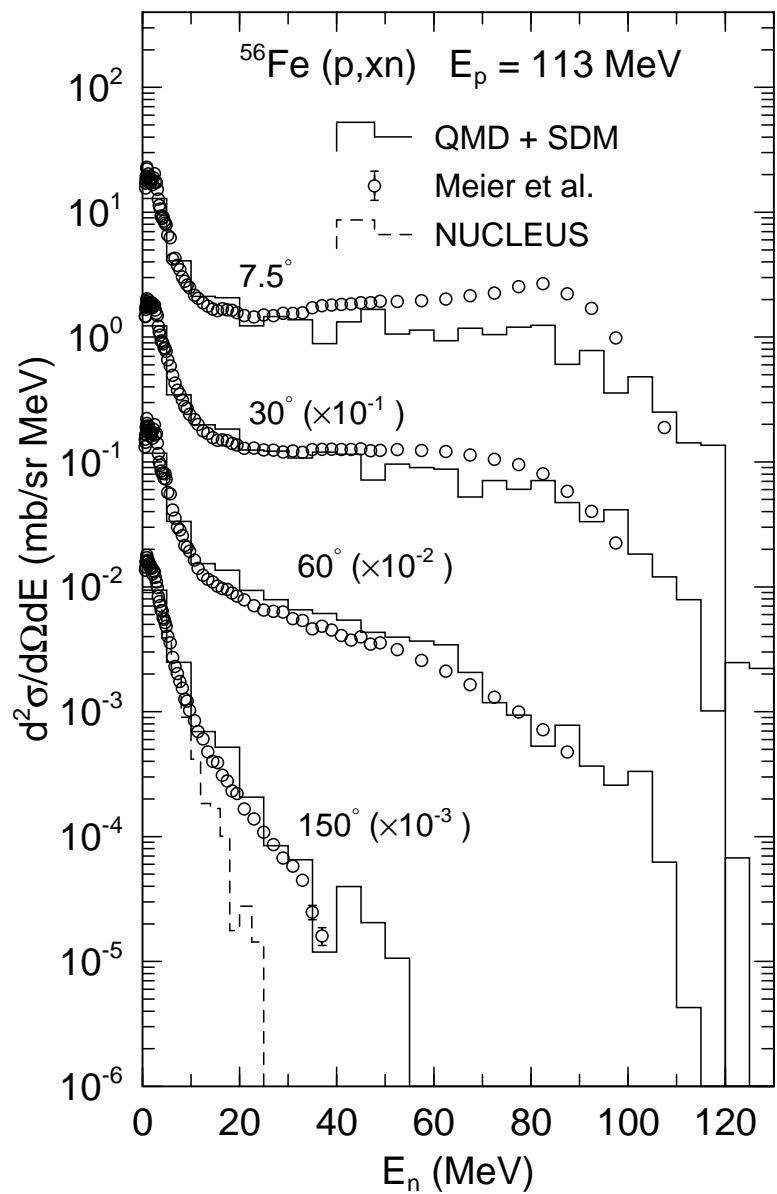
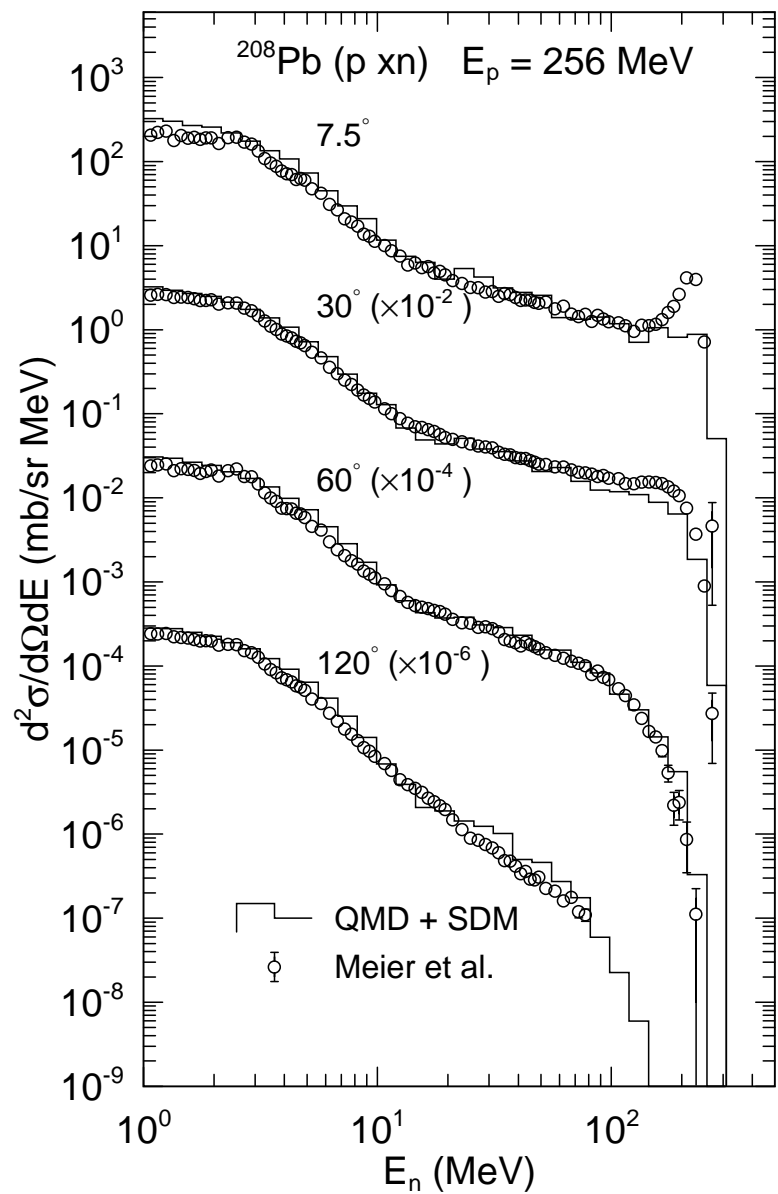
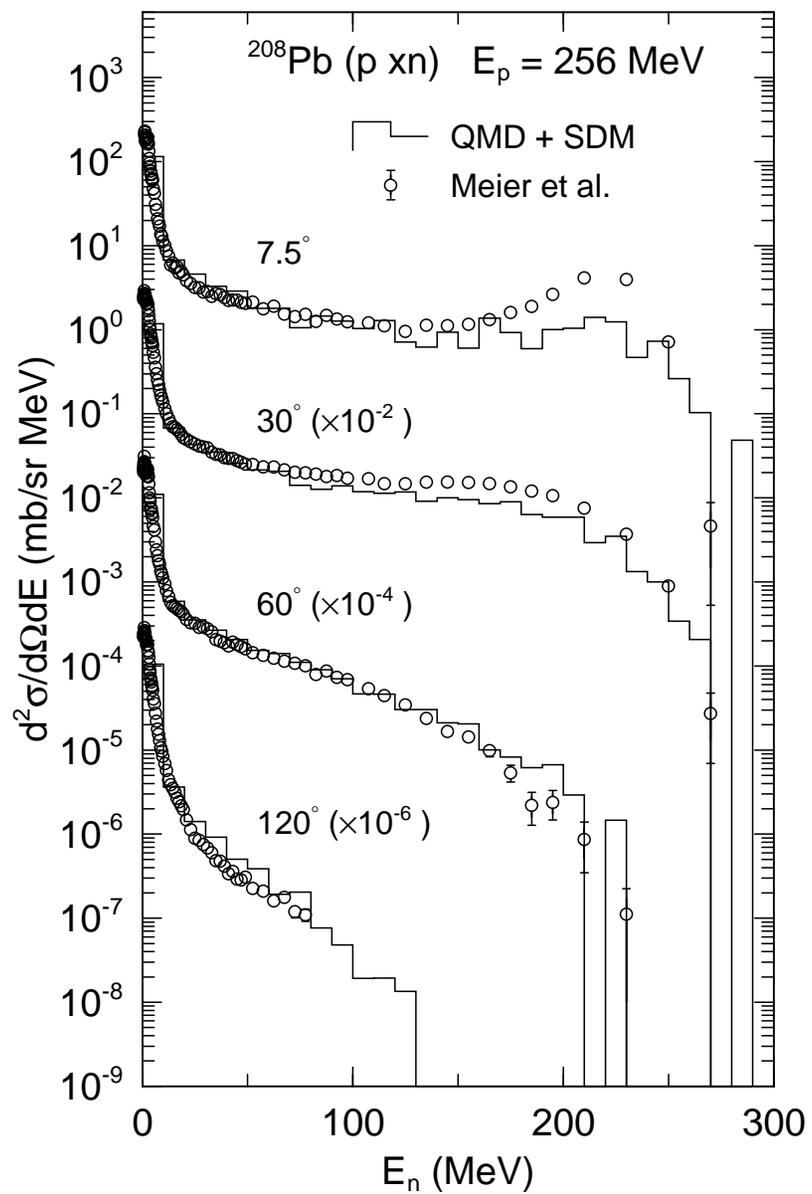
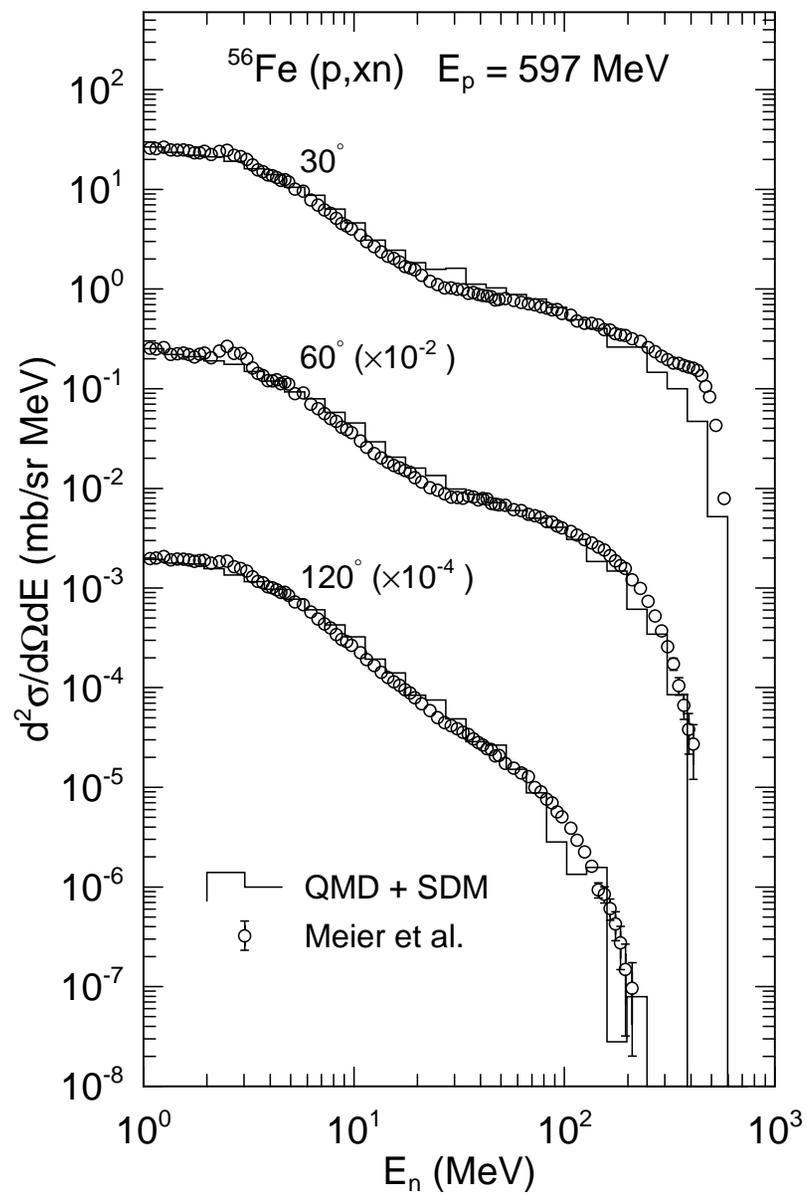
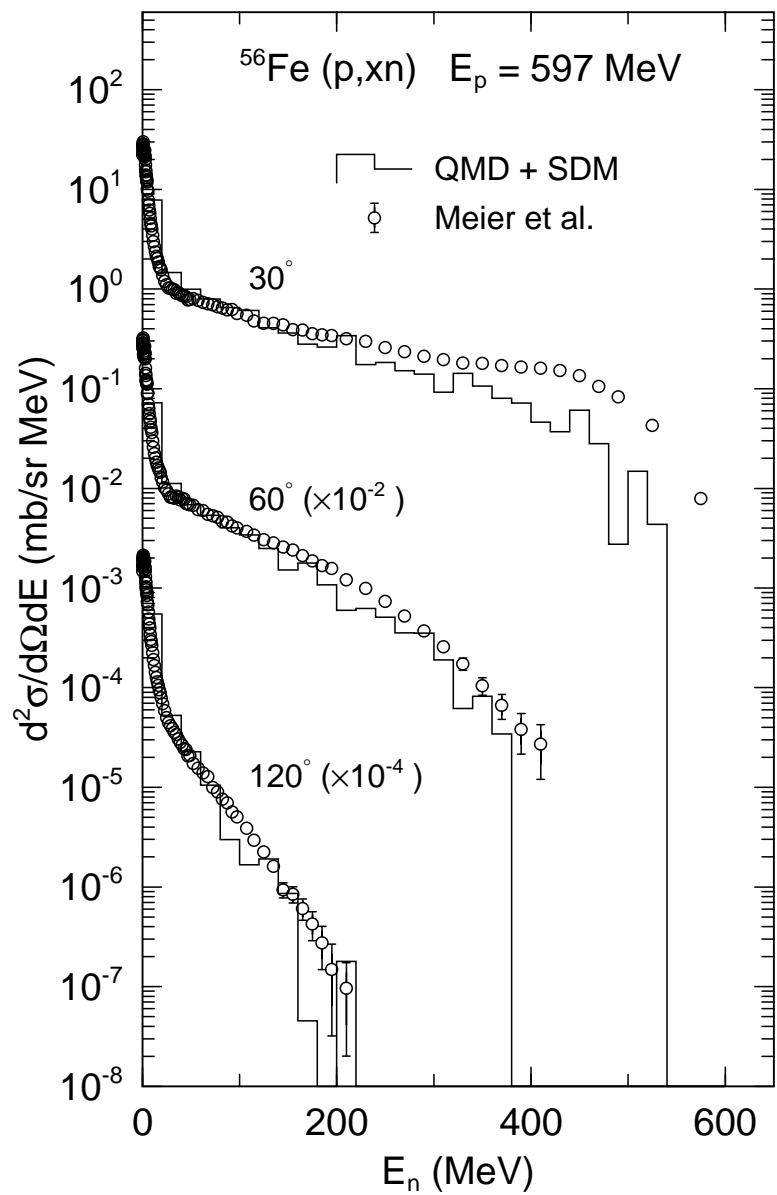
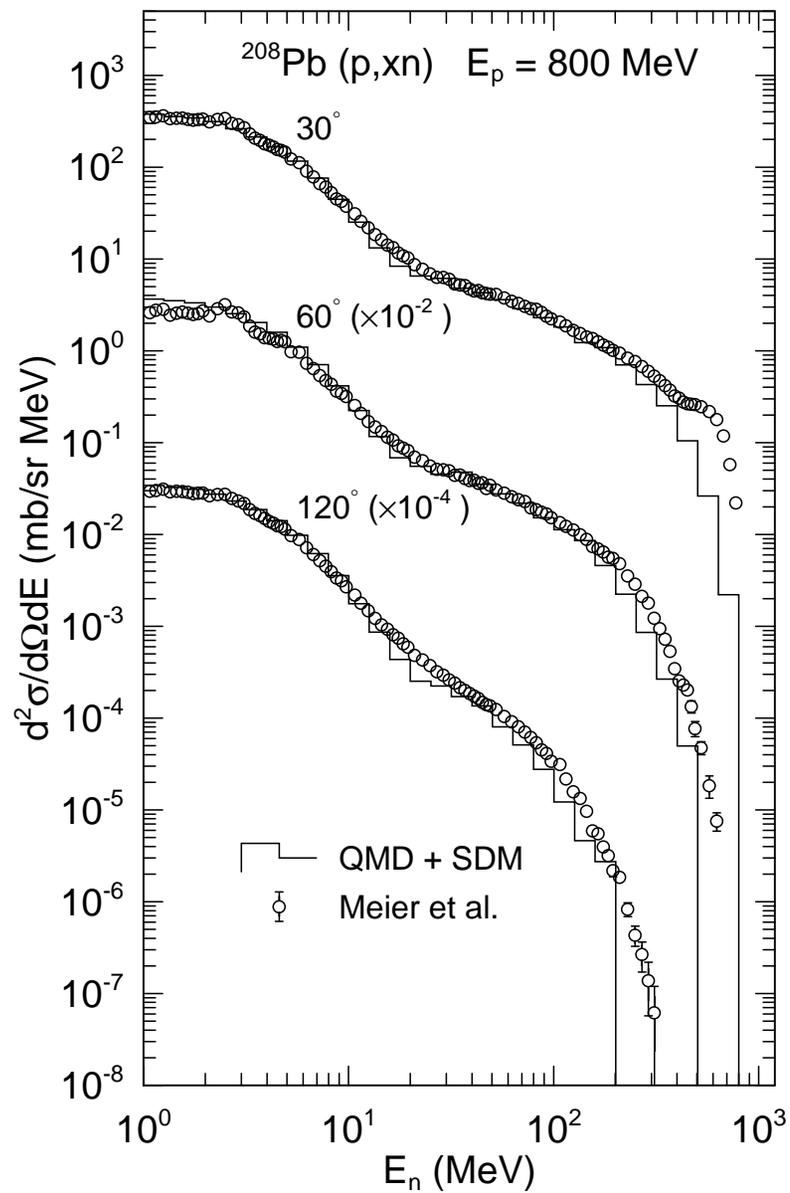
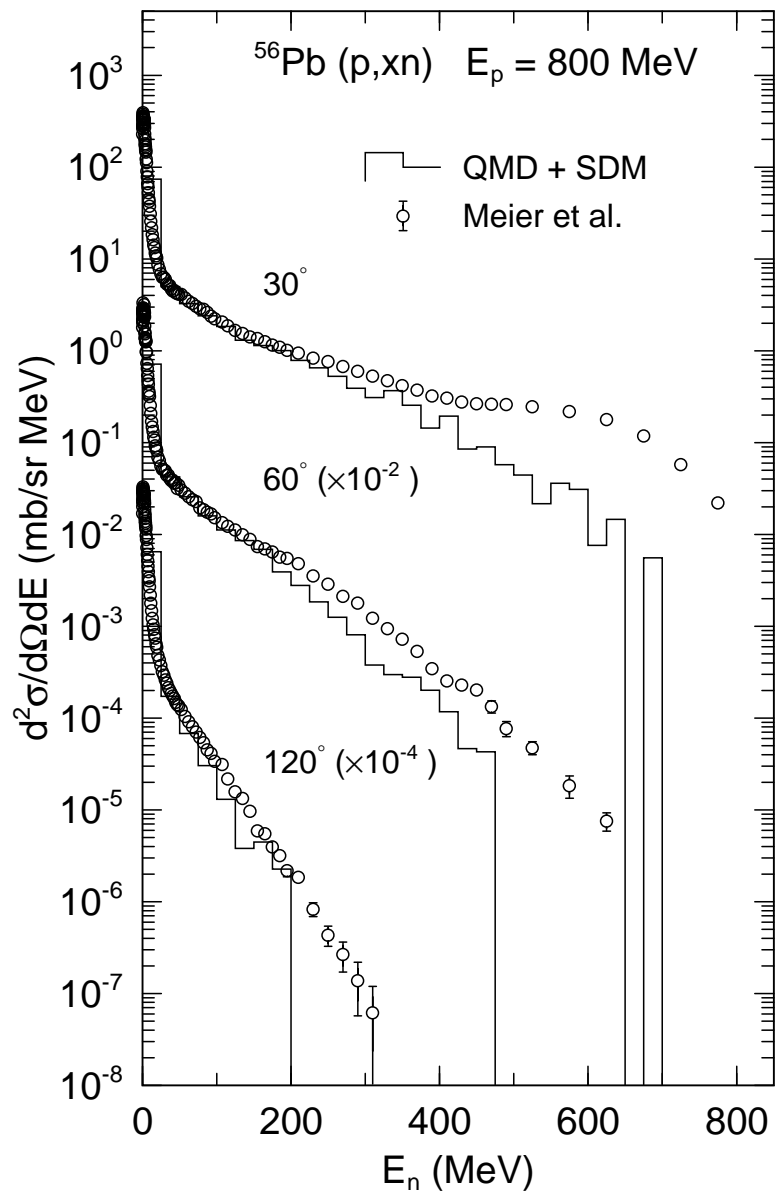
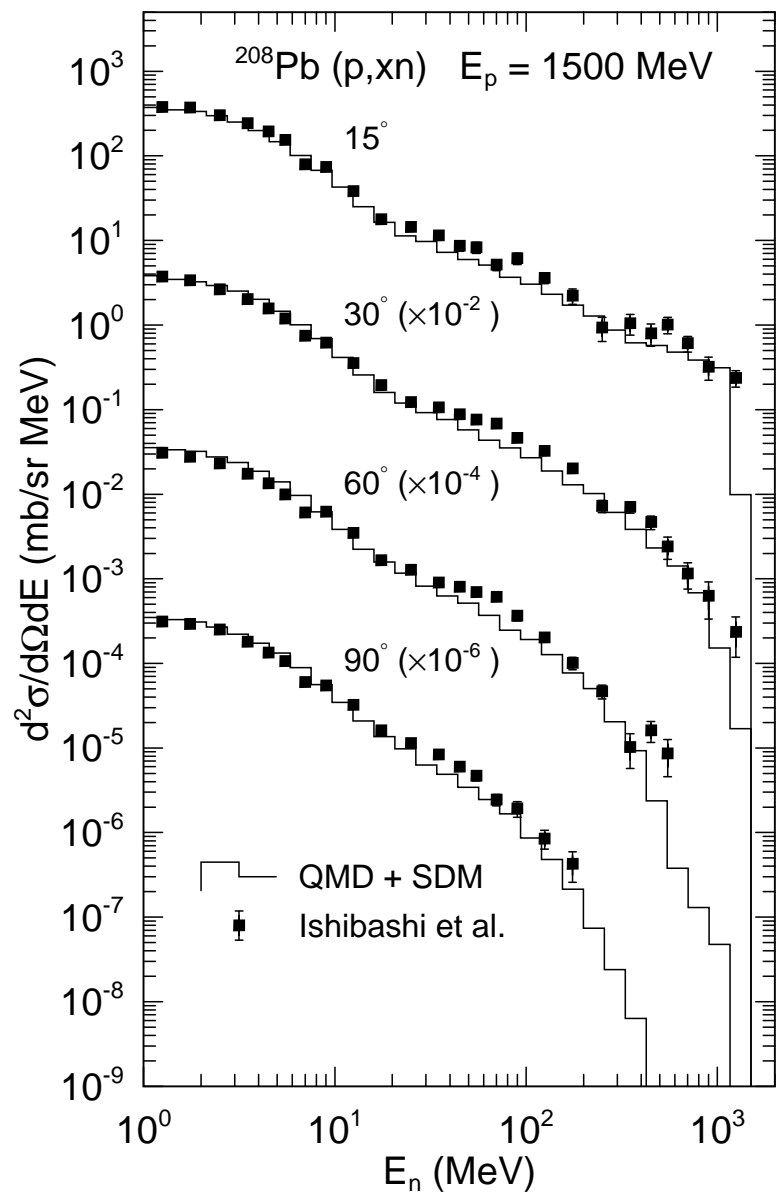
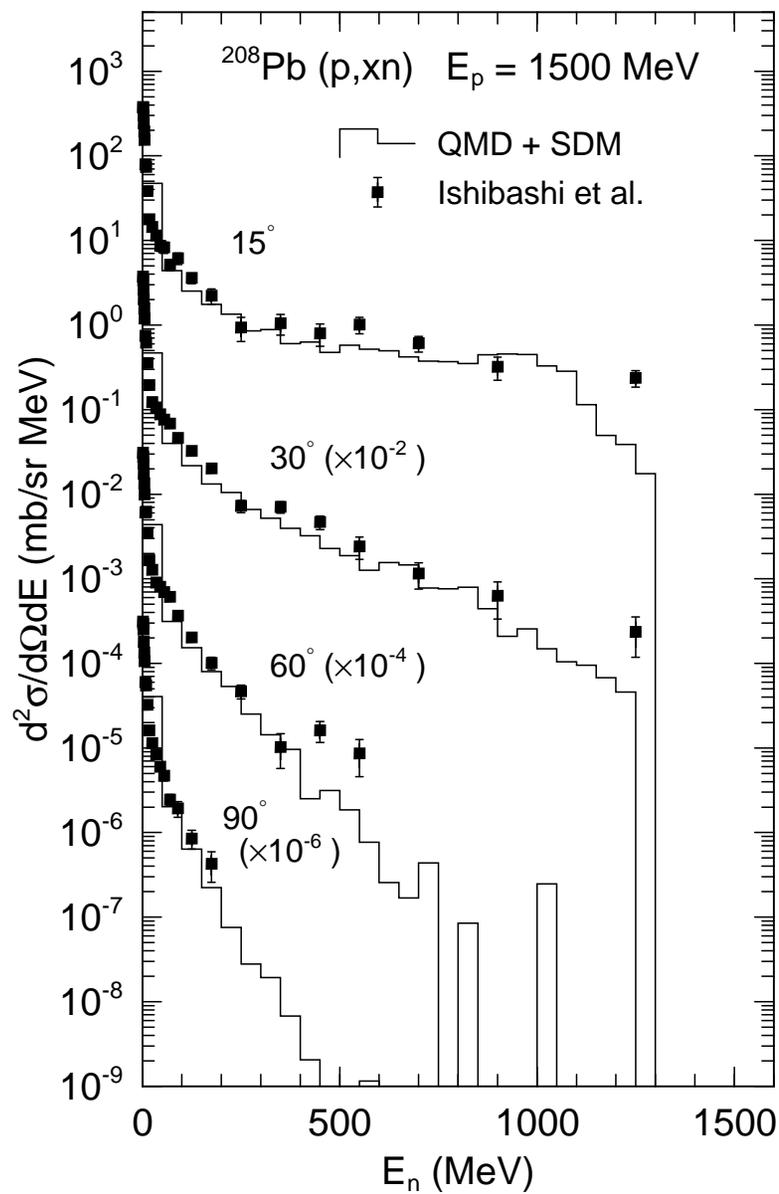
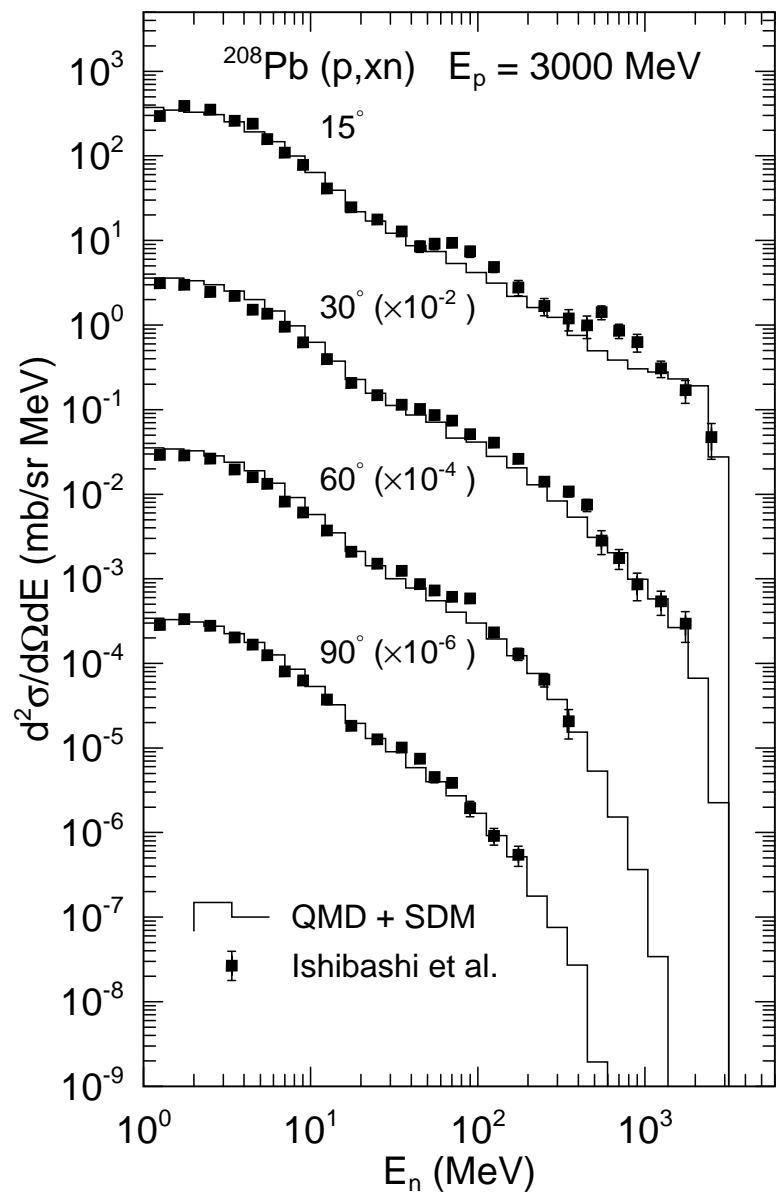
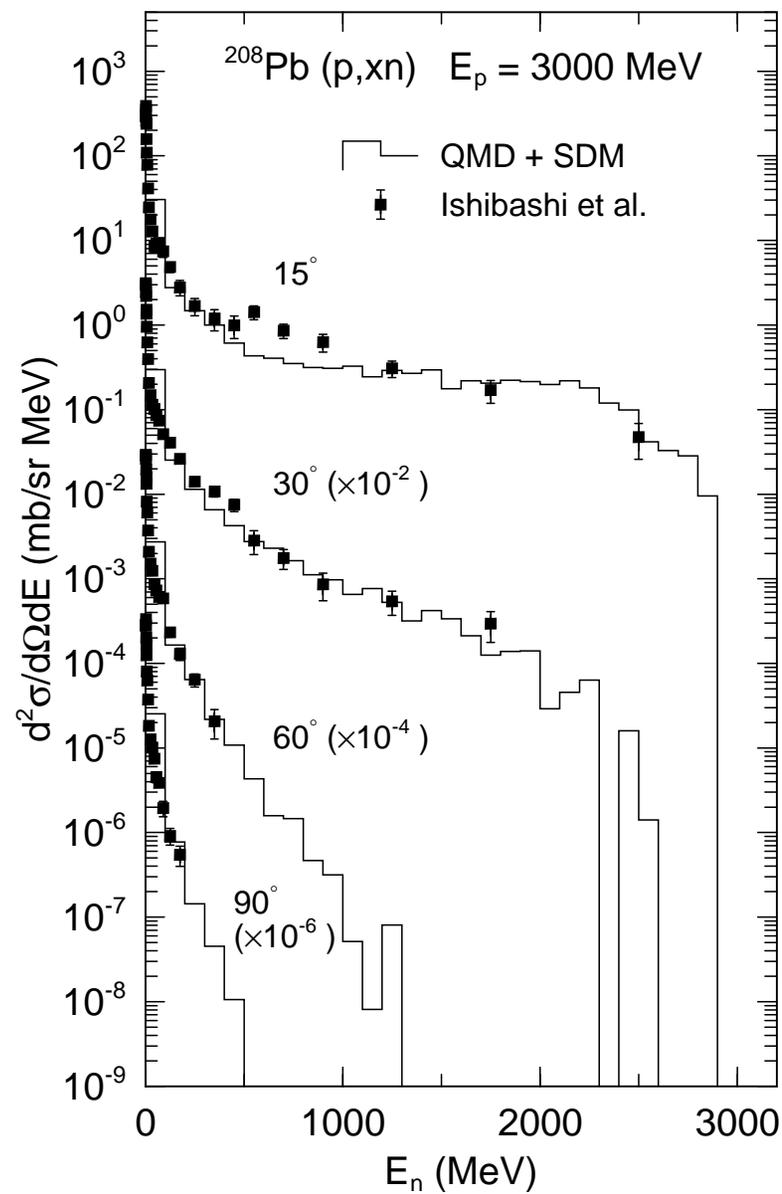
\end{document}